\documentclass[onecolumn]{aastex}
\usepackage{graphicx,amssymb,natbib}%,lscape}
\usepackage{amsmath}

\newcommand{\kms}{km s$^{-1}$}

\newcommand{\etal}{et al.}

\newcommand{\degrees}{\ensuremath{^\circ}}

\newcommand{\Halpha}{H$\alpha$}
\newcommand{\HI}{\mbox{H\textsc{i}}}

\newcommand{\solarmassespersquareparsec}{\ensuremath{M_\odot \text{pc}^{-2}}}

\newcommand{\QN}{\ensuremath{\mathcal{Q}_N}}

\newcommand{\Htwo}{H$_2$}
\newcommand{\per}[1]{\ensuremath{^{-#1}}}

\shortauthors{Hallenbeck \etal}

\begin{document}
\shorttitle{HIghMass Galaxies in \HI\ \& \Htwo.}
\title{HIghMass - High HI Mass, HI-rich Galaxies at z$\sim0$:\\Combined HI and \Htwo\ Observations}
\author{Gregory Hallenbeck\altaffilmark{1}, Shan Huang\altaffilmark{2}, Kristine Spekkens\altaffilmark{3}, Martha P. Haynes\altaffilmark{4}, Riccardo Giovanelli\altaffilmark{4}, Elizabeth A. K. Adams\altaffilmark{5}, Jarle Brinchmann\altaffilmark{6}, John Carpenter \altaffilmark{7}, Jayaram Chengalur\altaffilmark{8}, Leslie K. Hunt\altaffilmark{9}, Karen L. Masters\altaffilmark{10,11}, and Am\'{e}lie Saintonge\altaffilmark{12}}

\altaffiltext{1}{Union College, Department of Physics \& Astronomy; hallenbg@union.edu}
\altaffiltext{2}{CCPP, New York University, 4 Washington Place, New York, NY 10003; shan.huang@nyu.edu}
\altaffiltext{3}{Royal Military College of Canada, Department of Physics, PO Box 17000, Station Forces, Kingston, Ontario, Canada K7K 7B4; Kristine.Spekkens@rmc.ca}
\altaffiltext{4}{Cornell Center for Astrophysics and Planetary Science (CCAPS), Space Sciences Building, Cornell University, Ithaca, NY 14853; haynes@astro.cornell.edu, riccardo@astro.cornell.edu}
\altaffiltext{5}{ASTRON, the Netherlands Institute for Radio Astronomy, Postbus 2, 7990 AA, Dwingeloo, The Netherlands; adams@astron.nl}
\altaffiltext{6}{Leiden Observatory, Leiden University, P.O. Box 9513, 2300 RA Leiden, Netherlands {\textit{e-mail:}} jarle@strw.leidenuniv.nl}
\altaffiltext{7}{California Institute of Technology Dept. of Astronomy, MC 249-17, Pasadena, CA 91125; jmc@astro.caltech.edu}
\altaffiltext{8}{National Centre for Radio Astrophysics, Tata Institute for Fundamental Research, Pune 411 007, India; chengalu@ncra.tifr.res.in}
\altaffiltext{9}{INAF-Osservatorio Astrofisico di Arcetri, Largo E. Fermi 5, I-50125, Firenze, Italy; hunt@arcetri.inaf.it}
\altaffiltext{10}{Institute of Cosmology and Gravitation, Dennis Sciama Building, Burnaby Road, Portsmouth POI 3FX; Karen.Masters@port.ac.uk}
\altaffiltext{11}{South East Physics Network, www.sepnet.ac.uk}
\altaffiltext{12}{University College London Dept. of Physics \& Astronomy, Kathleen Lonsdale Building, Gower Place, London, WC1E 6BT, United Kingdom; a.saintonge@ucl.ac.uk}

\begin{abstract}
\noindent
We present resolved \HI\ and CO observations of three galaxies from the HIghMass sample, a sample of \HI-massive ($M_\text{HI}>10^{10} M_\odot$), gas-rich ($M_\text{HI}$ in top 5\% for their $M_*$) galaxies identified in the ALFALFA survey. Despite their high gas fractions, these are not low surface brightness galaxies, and have typical specific star formation rates (SFR$/M_*$) for their stellar masses. The three galaxies have normal star formation rates for their \Htwo\ masses, but unusually short star formation efficiency scale lengths, indicating that the star formation bottleneck in these galaxies is in the conversion of \HI\ to \Htwo, not in converting \Htwo\ to stars. In addition, their dark matter spin parameters ($\lambda$) are above average, but not exceptionally high, suggesting that their star formation has been suppressed over cosmic time but are now becoming active, in agreement with prior \Halpha\ observations. %For at least one (UGC 6168), there is some evidence for noncircular motions in the gas disk, as was previously observed in HIghMass galaxy UGC 9037. Overall, these properties 

\end{abstract}

\section{Introduction}
\noindent
Star-forming galaxies in the local universe follow a tight correlation between specific star formation rate ($\text{SSFR}\equiv\text{SFR}/M_*$) and stellar mass known as the star-forming main sequence: as stellar mass increases, SSFR slowly decreases (e.g. \citealt{Brinchmann2004}, \citealt{Salim2007}, \citealt{Schiminovich2007}). This main sequence has been observed out to high stellar masses ($M_*>10^{10} M_\odot$) in the optically selected GASS survey (the GALEX Arecibo SDSS survey; \citealt{GASS}, \citealt{Schiminovich2010}). However, optically-selected samples are inherently biased towards galaxies with higher surface brightnesses. In comparison to optical samples, samples selected by \HI\ are bluer, have higher gas fractions ($\text{GF}\equiv M_\text{HI}/M_*$), lower star formation efficiencies ($\text{SFE}\equiv\text{SFR}/M_\text{HI}$), and lower surface brightness than optically-selected samples (e.g. \citealt{Huang2012a}). Given the strong differences between optically and \HI-selected samples, it not immediately clear whether galaxies with high \HI\ masses also follow the star formation main sequence.

Consider the ``\HI\ Monsters'' sample of \citet{Lee2014} and the Bluedisks sample of \citet{Bluedisks}, which were selected on the basis of \HI\ mass, either directly from the ALFALFA survey (in the case of the \HI\ Monsters), or as inferred from optical colors (Bluedisks), with each sample yielding \HI\ masses of $M_\text{HI}>10^{10.5} M_\odot$ and $10^{8.3} M_\odot < M_\text{HI}<10^{10.4} M_\odot$, respectively. The \HI\ Monsters have high \HI\ masses, and correspondingly high $M_*\sim10^{11}M_\odot$ and $M_{H_2}\sim10^{10}M_\odot$. They also have quite high star formation rates, and their SSFRs lie on the star forming main sequence. The Bluedisks galaxies' \HI\ radii follow scaling relations derived for lower $M_\text{HI}$ galaxies. Additionally, their \HI\ disks do not appear disturbed, suggesting that they had not recently acquired gas from a merger. It appears that selecting on the basis of \HI\ mass alone yields samples which are similar to lower mass spiral galaxies, but ``scaled up'' to higher total mass. As gas fraction decreases with increasing stellar mass, these samples consist of \HI-massive, but not particularly gas-rich galaxies. Instead, the typical gas fractions of both samples are in the range $0.1<\text{GF}<1.0$.

Many galaxies which are both \HI-massive ($M_\text{HI}>10^{10} M_\odot$) and gas-rich ($GF\gtrsim1$) deviate from the star-forming main sequence. Some are giant, low-surface brightness galaxies (GLSBs) like Malin~1 ($M_\text{HI}=10^{10.8}M_\odot$, $GF\approx0.9$; \citealt{Bothun1987}; \citealt{Lelli2010}). GLSBs are also seen to also exhibit low surface densities of \HI. \citet{Lemonias2014} examined an optically selected-sample of GASS galaxies with $M_\text{HI}>10^{10} M_\odot$ and high gas fractions. As a whole, the sample has suppressed star formation, lying in the same region of $M_*-SSFR$ space as GLSBs. Karl G. Jansky Very Large Array (VLA) observations of the sample show that these galaxies have extended, low average deprojected \HI\ surface density ($\Sigma_{\text{HI}}$), as well as low deprojected SFR surface density ($\Sigma_\text{SFR}$).

From the 40\% sky area data release of the ALFALFA survey ($\alpha.40$; \citealt{alpha.40}), we have identified the HIghMass sample, first presented in \citet{Huang2012a}. Like the sample of \citet{Lemonias2014}, the HIghMass galaxies are selected to have both a high \HI\ mass ($M_\text{HI}>10^{10}M_\odot$) and unusually high gas fractions for their stellar masses (GF is 1~$\sigma$ above average). This yields a sample with $\text{GF}\gtrsim0.4$; half have $M_\text{HI}>M_*$. Despite these high gas fractions, unlike the optically-selected sample of \citet{Lemonias2014}, the HIghMass galaxies do not have suppressed star formation: they fall along the $\text{SSFR}-M_*$ star forming main sequence.

The only sample with properties similar to HIghMass is \HI GHz \citep{HIGHz}. The \HI GHz galaxies are similarly massive ($M_\text{HI}>10^{10}M_\odot$) and gas-rich ($0.1<GF<2$), while also lying on or above the star-forming main sequence. The galaxies of the \HI GHz sample lie at $0.17\leq z\leq0.25$, and are massive galaxies which are still assembling their disks. The  HIghMass galaxies are at a redshift of $z<0.06$, suggesting that the HIghMass galaxies are the low redshift analogs of the \HI GHz sample. But how can massive gaseous reservoirs like those observed in the HIghMass galaxies survive to $z = 0$ in galaxies whose star formation isn't suppressed? Our hypotheses broadly fall into two categories. First, the galaxies have unusually high dark matter halo spin parameters, suppressing time-averaged star formation. Second, their cold gas has been recently acquired. 

%- Huang et al 2014 found:
%	- Active ongoing star formation (they're not ``crouching giants'')
%	- High overall SFR, but the SF is more extended (lower surface densities)
%	- Relatively poor star formation in the past
%	- Suggest high spin parameter, both in comparison with optically selected samples and the ALFALFA parent sample

The dark matter spin parameter is a dimensionless way to quantify the angular momentum of a dark matter halo, $\lambda\equiv J\left|E\right|^{1/2}G^{-1}M^{-5/2}$. Theoretically, high spin parameters are associated with bluer colors, lower optical surface brightness, and higher gas fractions (e.g., \citealt{Jimenez1998}; \citealt{MoMaoWhite1998}; \citealt{Boissier2000}; \citealt{Maccio2007}). Unfortunately, $\lambda$ is ultimately a parameter which is not directly observable. Despite this, several works have attempted to infer it for populations of galaxies based on optical (\citealt{Hernandez2007}; \citealt{CSHernandez2009}) and a combination of optical and \HI\ properties \citep{Huang2012a}, with results agreeing with theory. \citet{Huang2012a} found that the ALFALFA population as a whole have elevated values of $\lambda$ compared with an SDSS-selected sample \citet{Hernandez2007}. \citet{Huang2014} further suggest that the HIghMass galaxies have, on average, even higher values of $\lambda$ than the overall ALFALFA sample.

Recent gas can come from a ``galactic fountain'' effect, where supernovae have ionized and ejected gas (\citealt{FraternaliBinney2006}, 2008; \citealt{Oppenheimer2010}). While outside of the galactic disk, the gas is unable to form stars. Over time, the gas can cool, recombine and return to the disk. Studies of NGC 891 and NGC 2403 \citep{FraternaliBinney2008} at a distance of $<10$ Mpc have inferred reaccretion rates as high as $1-3$ $M_\odot$ yr$^{-1}$, similar to their star formation rates. Simulations by \citet{Marinacci2010} find that in most cases, this extraplanar gas is unlikely to be of high enough column density to be observed directly, but can contribute to a galaxy's global profile---that is, there can exist gas which is unable to contribute to star formation but will contribute to the \HI\ mass observed by ALFALFA. Such gas shares a common specific angular momentum with the existing gaseous disk, but can cause inflows \citep{Fraternali2001}. Alternatively, recently acquired gas may originate in the intergalactic medium, having been unassociated with any galaxy until now. Such gas has angular momentum which is uncorrelated with the galaxy's disk, leading to warps as well as flows \citep{FraternaliBinney2008}.

Previous work by \citet{Hallenbeck2014} examined in detail two of the HIghMass galaxies using $\sim3$ kpc resolution VLA observations. The two galaxies have quite similar optical photometric and unresolved \HI\ spectral  properties: $M_*>10^{10} M_\odot$, $M_\text{HI}>10^{10.3}$, specifically with $M_\text{HI}>M_*$. However, upon resolving the \HI, we see that the two galaxies are drastically different. One (UGC 12506) is a low surface brightness (LSB) galaxy with a very high dark matter halo spin parameter ($\lambda=0.15$), low surface density, extended \HI\ (typically $1-5$\solarmassespersquareparsec\ at radii from $10-40$ kpc), and low star formation surface densities. The other (UGC 9037) has an elevated high spin parameter ($\lambda=0.07$), but has both high central \HI\ surface density ($>10\solarmassespersquareparsec$ at radii less than 10 kpc) and centrally peaked star formation. In addition, UGC 9037 has what appear to be high-velocity inflows at all radii, with a peak non-circular $v = 0.09v_\text{rotation}$, suggesting that the galaxy is undergoing transition to a more intense star-forming phase.

% WHAT'S GOING ON IN THIS PAPER

This paper is the second in a series, building on the results of \citet{Hallenbeck2014}. We present observations of the $^{12}$CO(1-0) line (as a proxy for H$_2$), observed using CARMA (the Combined Array for Research in Millimeter-wavelength Astronomy) for three galaxies, UGC 6168, UGC 7899, and NGC 5230 (which is also UGC 8573). These three galaxies were specifically selected for study because they had the highest predicted CO column density using the scaling relations derived by \citet{COLD-GASS}. These new observations are combined with the \Halpha\ studies of \citet{Huang2014} to examine disk stability and the star formation efficiency of the \HI\ and \Htwo\ phases and determine whether the star formation bottleneck is in the conversion of \HI\ to \Htwo, or \Htwo to stars. In addition, in order to test the recent accretion hypothesis, we are studying the resolved gas velocity fields to search for gas inflows and warps. Finally, we directly derive values of $\lambda$ for each galaxy.

Much of our data reduction mirrors that in the previous work of \citet{Hallenbeck2014}. We summarize and discuss differences from that previous work in \S\ref{sec:data}. Results for each individual galaxy are presented in \S\ref{sec:results}. We discuss possible evolutionary histories for each galaxy as well as the HIghMass sample in general in \S\ref{sec:discussion}. Our conclusions are summarized in \S\ref{sec:conclusions}.

\section{Observations and Data Reduction}
\label{sec:data}
\noindent
In Table 1 we present the global gas, stellar, and star formation properties of the three HIghMass galaxies studied in this work, as well as the two galaxies studied by \citet{Hallenbeck2014}. Except as described in the following sections, data reduction and analysis methods for gas are identical to those in \citet{Hallenbeck2014}; methods for deriving stellar and star formation properties follow the methods of \citet{Huang2014}. A brief summary of those methods are as follows:
\begin{itemize}
 \item Total \HI\ masses, recessional velocities, and distances are taken from the $\alpha.40$ catalog \citep{alpha.40}, which assumes $H_o = 70$ \kms\ Mpc$^{-1}$.
 \item \HI\ observations were performed at the VLA and GMRT; data reduction uses a combination of standard CASA and GIPSY packages. These methods recover the total \HI\ flux of our galaxies, and produce line profiles which agree with the single dish $\alpha.40$ catalog.
 \item Rotation curves are derived by fitting tilted rings to the observed moment 1 velocity fields using the GIPSY task \textsc{rotcur}, with radii spaced every half beam width. Non-circular velocities are then re-examined and confirmed using the DiskFit package (\citealt{SpekkensSellwood2007}; \citealt{SellwoodSanchez2010}; \citealt{KuziodeNaray2012}). %DiskFit uses bootstrap resampling methods to estimate the true uncertainties of the non-circular velocities; \textsc{rotcur} tends to underestimate uncertainties by a factor of $2-3$; the increased uncertainties are not reflected in the figures.
 \item Star formation rates and surface densities are calculated from \Halpha\ imaging taken at Kitt Peak National Observatory (KPNO). We take the resulting profiles directly from \citet{Huang2014}.
 \item Dark matter properties are determined by fitting to the observed rotation curves using the GIPSY task \textsc{rotmas}. We model each galaxy as having thin gas (\HI, \Htwo, and He) and stellar components along with dark matter.
 \item A modified dark matter halo spin parameter $\lambda^\prime$ is calculated directly from the resolved gas properties and the dark matter halo fit (Equations 4, 5, and 6 from \citealt{Hallenbeck2014}):
 \begin{equation}
  \lambda^\prime = \frac{J_\text{HI}|E|^{1/2}}{GM^{5/2}} = \frac{1}{2}\frac{\sum_{i}\frac{M_{\text{HI},i}}{M_\text{HI}}V_ir_i \cdot V^2_C}{GM^{1/2}}
 \end{equation}
 Where $M_{\text{HI},i}$, $V_i$, and $r_i$ are the \HI\ mass, velocity, and galactocentric radius of each tilted ring, $V_C$ is the maximum circular velocity of a fit pseudo-isothermal halo, and $M$ is the mass of the dark matter halo. $M$ is obtained via abundance matching both the combined stellar and \HI\ masses \citep{baryonmassfunction}.
\end{itemize}
\noindent
%All three galaxies studied in this work exhibit significant non-circular velocities at a wide range of radii. Because tilted ring fits do not have a unique solution and non-circular velocities are degenerate with position angle (PA) and inclination ($i$), it is important that our final tilted ring fit results be as physically motivated as possible. The inner disks of real galaxies all appear to be quite flat, with nearly constant PA and $i$. To that end we first determine a best-fit average value of both PA and $i$, then produce rotation curves holding both constant.
%\begin{centering}
\begin{deluxetable}{cccccccc}
\newpage
\tablecolumns{7}
\tablewidth{0pt}
\tabletypesize{\scriptsize}
\tablecaption{Sample Optical and Radio Properties}
\tablehead{
 \colhead{Galaxy} & \colhead{d} & \colhead{$\log M_\text{HI}$} & \colhead{$\log M_{\text{H}_2}$} & \colhead {$R_{25}$} & \colhead{$\log M_*$ (SED)} & \colhead{$\log M_*$ (IRAC)} & \colhead{log SFR}   \\
                  & Mpc         & $M_\odot$                    & $M_\odot$                       & kpc                 & $M_\odot$                  & $M_\odot$                   & $M_\odot$ yr\per{1} \\
							(1) & (2)         & (3)                          & (4)                             & (5)                 & (6)                        & (7)                         & (8)}
\startdata
UGC 6168          & 120         & 10.35                        & \ 8.96                          & 23.8 & 10.37                      & 10.59                       & 0.57                \\
UGC 7899          & 128         & 10.42                        & \ 9.68                          & 36.8 & 10.49                      & 10.93                       & 1.20                \\
NGC 5230          & 101         & 10.53                        & 10.02                           & 36.3 & 10.89                      & 11.22                       & 0.96                \\
\hline
UGC 9037          & \ 88        & 10.33                        & ---                             & 23.0 & 10.09                      & ---                         & 0.56                \\
UGC 12506         & \ 98        & 10.53                        & ---                             & 40.0 & 10.46                      & ---                         & \ 0.40$^*$               
\enddata
\tablecomments{Optical and radio properties of the HIghMass galaxies in this work and in \citet{Hallenbeck2014}. Column 1: galaxy identifier; Column 2: Hubble flow distance of galaxy, from \citet{alpha.40}; Column 3: \HI\ mass, from \citet{alpha.40}; Column 4: Inferred \Htwo\ mass from CARMA observations of $^{12}$CO($J=1-0$) line; Column 5: Radius of the 25 mag arcsec$^{-2}$ isophote in $r$-band; Column 6: stellar mass, derived from fitting SEDs to SDSS magnitudes, from \citet{Huang2012a}; Column 7: stellar mass, derived from Spitzer observations, and the mass-to-light ratios of \citet{Querejeta2014}; Column 8: \Halpha-derived star formation rates, from \citet{Huang2014} ($^*$the SFR of UGC 12506 is an exception and is derived from SED fitting its SDSS magnitudes).\label{tab:sample}}
\end{deluxetable}
%\end{centering}

\subsection{CO Observations \& Inferred H$_2$ Properties}
\label{sec:COdata}
\noindent
The three HIghMass galaxies were observed using CARMA in its compact E configuration. The CARMA visibilities for the three galaxies are exported from the native MIRIAD format into CASA before following the same data reduction techniques as were performed for the \HI\ observations. The VLA has a primary beam full width half haximum of $\sim30\arcmin$ when observing the 21 cm \HI\ line. However, at 115 GHz, the CARMA primary beam scale is $\sim1\arcmin$, similar to the sizes of our galaxies (major axes $\sim1\arcmin$). We image out to the 20\% response contour, and correct the fluxes accordingly.%For all three cases, the 20\% contour contains all or essentially all of the CO emission above noise thresholds.

Single-dish $^{12}$CO(1-0) observations of 18 HIghMass galaxies were performed with the 30m IRAM telescope, including UGC 6168, UGC 7899, and NGC 5230. For all three galaxies, we produced spectra from the CARMA data cubes corresponding to the the 22$^{\prime\prime}$ field of view of the 30m IRAM dish. The fluxes derived from these spectra agree with the fluxes derived from the single-dish observations (Huang et al 2017 in preparation).

We must assume a conversion factor between the observed CO luminosity $L^\prime_\text{CO}$ (in units of K \kms\ pc$^{2}$) and the total H$_2$ mass, known as $\alpha_\text{CO}$. The conversion factor is known to vary with metallicity (e.g. \citealt{Wilson1995}, \citealt{Arimoto1996}, \citealt{Bolatto2013}, \citealt{Sandstrom2013}). However, even though the HIghMass galaxies are gas-dominated, they all have $M_*\gtrsim10^{10}$ M$_\odot$, and thus we expect them to have metallicities similar to the Milky Way. We follow \citet{COLD-GASS} and use a Milky Way value for $\alpha_\text{CO}$ averaged over several recent measurements (\citealt{StrongMattox1996}; \citealt{DameHartmannThaddeus2001}; \citealt{Blitz2007}; \citealt{Draine2007}; \citealt{Heyer2009}; \citealt{Abdo2010}), $\alpha_\text{CO}=3.2 $M$_\odot ($K \kms\ pc$^{2})^{-1}$. Given the conversion factor, we then compute the total H$_2$ mass following \citet{Solomon1997}:
\begin{equation}
  \label{eqn:MH2}
  M_{\text{H}_2} = L^\prime_{\text{CO}} \alpha_\text{CO} = 3.25\times10^{7} \alpha_\text{CO} S_\text{CO} \nu^{-2} d^2
\end{equation}
where $S_\text{CO}$ is the total integrated CO flux in Jy \kms\ (which are our intensity map units) and $\nu$ is the rest frequency of the line in GHz, which for $^{12}$CO(1-0) is 115.271 GHz, and $d$ is the Hubble flow distance of the galaxy in Mpc.

Production of \Htwo\ moment maps, fitting of rotation curves, and deriving deprojected surface densities then follows the same method as for \HI. However, in most cases, the CO is only partially resolved by CARMA, and so we must assume inclinations and position angles as derived by the \HI\ rotation curves.
%With this conversion made, the deprojected surface densities of H$_2$ can be calculated in a manner analogous to that of \HI, depending on whether the galaxy is nearly face-on, inclined, or nearly edge-on.

\subsection{Stellar Masses}
\label{sec:stellarmass}
\noindent
\citet{Hallenbeck2014} used a combination of two methods to calculate stellar masses. Global stellar masses were calculated based on fitting model spectral energy distributions to SDSS magnitudes (\citealt{Salim2007}; \citealt{Huang2012b}; \citealt{Huang2012a}). These were then used to constrain the less precise surface density profiles following the method of \citet{massmodels}.

Here, instead of relying on optical photometry, we use infrared 3.6\micron\ and 4.5\micron\ photometry taken with Spitzer IRAC (\citealt{Spitzer}; \citealt{IRAC}), which should trace the old stellar population. We then convert to stellar masses following Equation (4) of \citet{Querejeta2014}:
\begin{equation}
 \log \Upsilon_{3.6} = -0.339([3.6] - [4.5]) - 0.336
\end{equation}
where $\Upsilon_{3.6}$ is the mass to light ratio of the 3.6\micron\ band, and $[3.6]-[4.5]$ is the color from the 3.6\micron\ and 4.5\micron\ bands. For UGC 6168, UGC 7899, and NGC 5230, the IRAC stellar masses are larger by a factor of $2-3$ compared with the SED fit masses. It is possible that this method is overestimating our stellar masses because the IRAC bands may have hot dust contamination as a result of dust heating by active star formation. %Later discussion in \S\ref{DMprops} shows that these Spitzer-derived total stellar masses yield dark matter halo fits with unphysical properties (i.e. $c<0$).

\subsection{Stability Criteria}

When calculating the stability of the disk via the Toomre $Q$ parameter, the work of \citet{Hallenbeck2014} neglected the contribution from stars. For both galaxies, the stellar mass was quite centrally concentrated by comparison with the \HI. In addition, $M_* < M_\text{HI}$. As a consequence, simple two-phase models of the galaxy disk, such as those by \citet{WangSilk1994} and \citet{Rafikov2001} predicted essentially no change in $Q$ at all radii.

However, the inclusion of \Htwo\ requires serious consideration. As \Htwo\ is much cooler and has a higher per-particle mass than \HI, its turbulent velocity is lower. As a consequence, for the same gas surface density, \Htwo\ is less stable. Therefore, in this paper we also consider the method of \citet{RomeoWiegert2011} and \citet{RomeoFalstad2013}. Their \QN\ accounts for an arbitrary number of phases. It is defined as (see Equation 19 of \citealt{RomeoFalstad2013}):

\begin{equation}
\label{eqn:QN}
\frac{1}{\mathcal{Q}_N} = \sum\limits_i\frac{\pi G \Sigma_i}{\sigma_i \kappa(r)}\frac{W_i}{T_i}
\end{equation}
The first factor is the Toomre $Q$ parameter for each particular phase, consisting of Newton's gravitational constant $G$, the surface density of the phase $\Sigma_i$, the velocity dispersion of the phase $\sigma_i$ and the epicyclic frequency $\kappa(r)$ calculated from the rotation curve. $W_i$ and $T_i$ are dimensionless parameters of order unity; these account for the overall stability being dominated by the most stable phase and the effect of the finite thickness of the disk, respectively (see Eqns. 16 and 18 from \citealt{RomeoFalstad2013}). Following \citealt{RomeoFalstad2013}, we use $\sigma_{\text{H}_2} \approx 6$ \kms\ \citep{Wilson2011} and $\sigma_\text{HI} \approx 11$ \kms\ \citep{Leroy2008}.

In the rest of this work, when we refer to ``Toomre $Q$'', we mean that we are treating all of the gas (\HI, \Htwo, and He) as a single phase with $\sigma_\text{gas}=11$ \kms, with He included as a factor of 1.33 correction to the mass. This will accurately model the \HI, but will overestimate the stability of the \Htwo\ from an artificially high $\sigma_\text{H$_2$}$. ``\QN'' refers to the multi-phase stability criteria of \citet{RomeoFalstad2013}, which properly accounts for the cold \Htwo.

\section{Results}
\label{sec:results}
\subsection{Overview of UGC 6168}
\begin{figure}
 \begin{center}
   \epsscale{0.9}
   \plotone{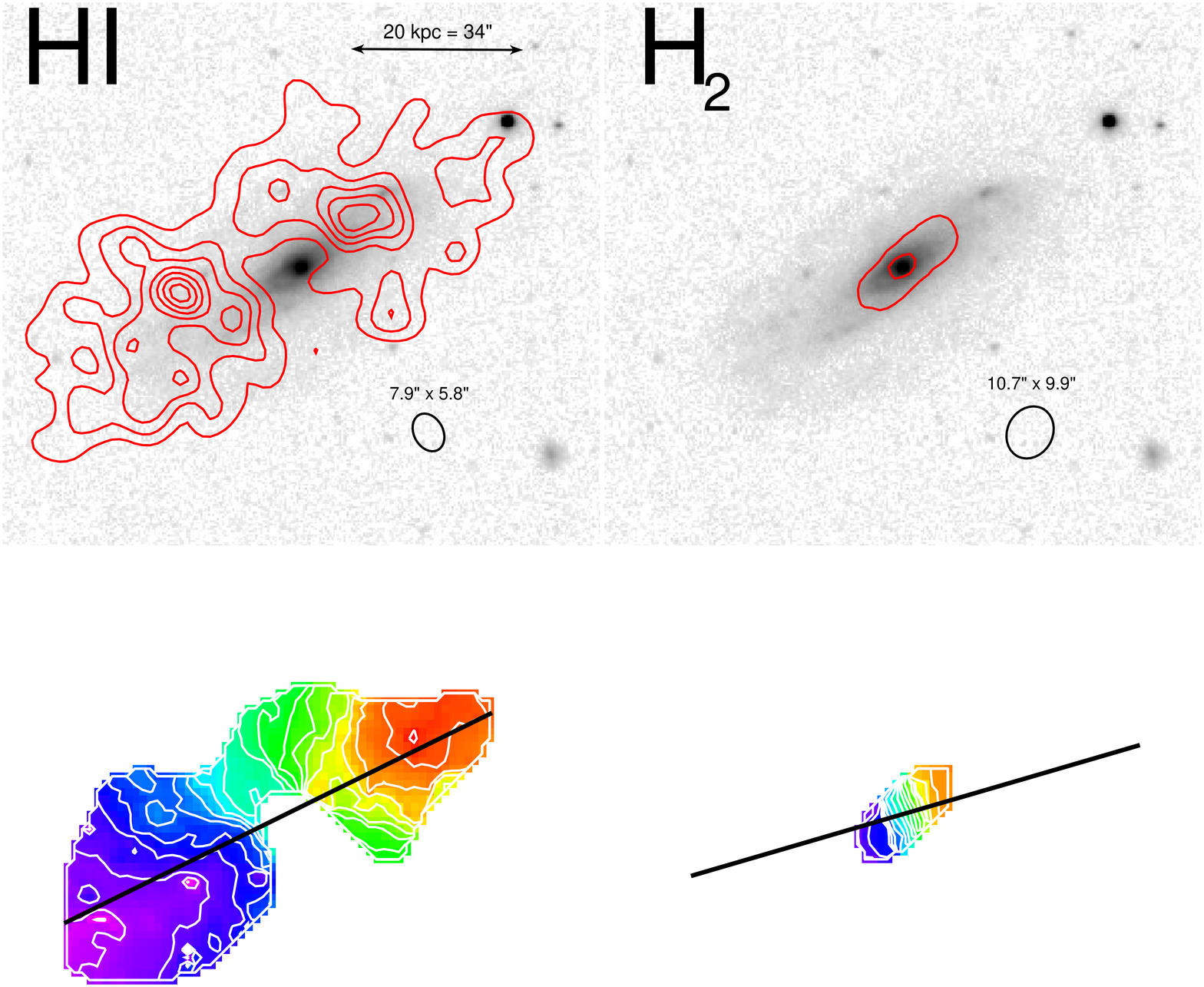}
 \end{center}
 \caption{(top left) \HI\ and (top right) CO integrated flux maps of UGC 6168. Contours begin at and are spaced every 5 \solarmassespersquareparsec\ of either \HI\ or inferred \Htwo. Both contours are overlaid on an inverted SDSS $r$ band image. (bottom left) \HI\ and (bottom right) CO velocity fields, with isovelocity contours spaced every 20 \kms. The \HI\ is depleted in the center, and appears in the shape of a ring. While the CO is just barely resolved, the major axis of rotation for the \HI\ and CO are different (black lines), which suggests the presence of non-circular flows in the galaxy.
 \label{fig:images-UGC6168}}
\end{figure}

\noindent
In Figure \ref{fig:images-UGC6168} (top), we present the integrated flux maps for our \HI\ and CO synthesis observations of UGC 6168. The contours begin at and are spaced every 5 \solarmassespersquareparsec\ of \HI\ or inferred \Htwo\ mass, as projected onto the sky. These contours are then overlaid on SDSS $r$-band images. The top left panel shows the \HI\ integrated flux map. Like the other two galaxies in this work, UGC 6168 is a spiral galaxy. \citet{Huang2014} note that the color gradient in UGC 6168 reverses, that is, as radius increases, the average color of the galaxy first becomes bluer, then redder in the outer disk. Color inversions may be related to declining star formation, dust-obscured central star formation, and outward migration of stars. This color gradient reversal is also a feature of UGC 7899 and NGC 5230.

The \HI\ is extended beyond the optical radius, as is typical for a gas-rich galaxy. In addition, the \HI\ within 5 kpc of the galaxy's center is clearly depleted, leaving a hole. This is unsurprising: the galaxies discussed in this work were specifically observed because they had the highest expected \Htwo\ masses out of the HIghMass sample. They are thus likely to be some of the most efficient at transforming gas into new stars. The top right panel shows the \Htwo\ distribution, as inferred from the CO emission. It very neatly fills in the hole left by the \HI.

The bottom panels of Figure \ref{fig:images-UGC6168} show the velocity fields of the \HI\ (left) and CO (right). Isovelocity contours are spaced every 20 \kms. %The southern half of the \HI\ velocity field shows . There is no corresponding contour on the north side of the galaxy, suggesting that the galaxy's rotation is not symmetric.
The CO emission is not sufficiently resolved by our 10\arcsec\ clean beam, and so no rotation curve can be fit. We nonetheless fit an average position angle, and the best-fit for each phase is overlaid. We note that the rotational axis of each phase is clearly different, which is indicative of noncircular flows.

\begin{figure}
 \begin{center}
  \epsscale{1.0}
	\plotone{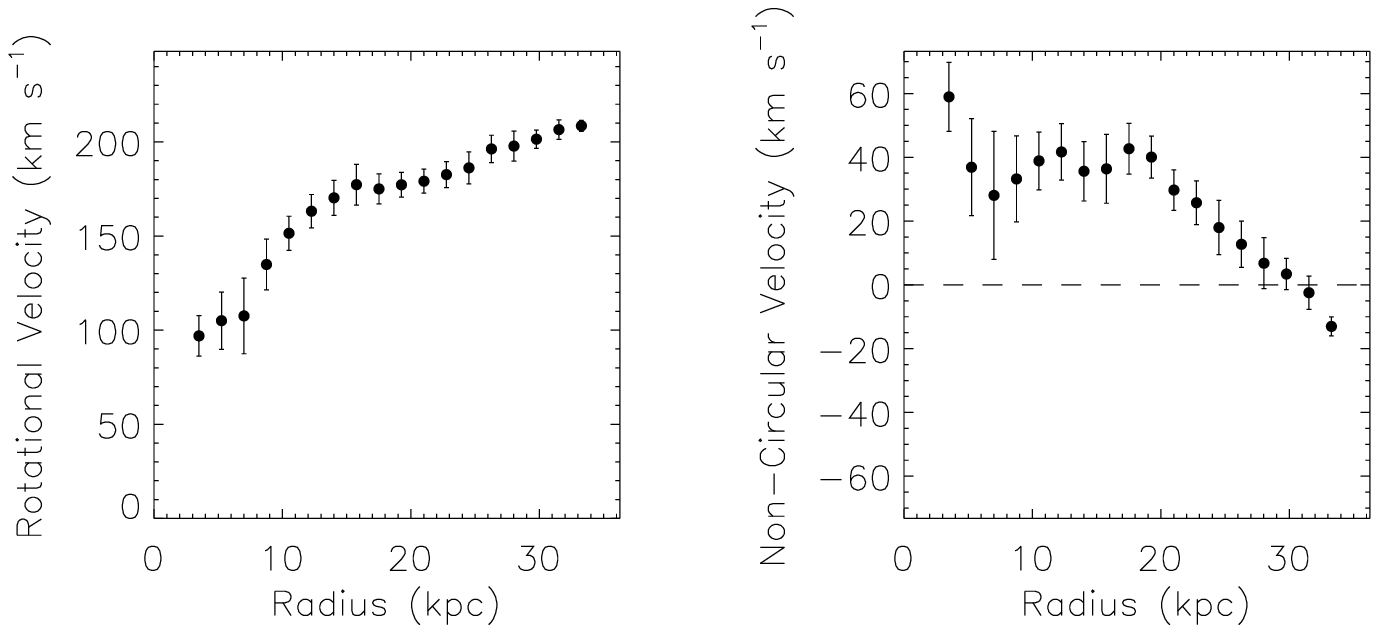}
 \end{center}
 \caption{Rotation curve of UGC 6168 derived by fitting a tilted ring model to the \HI\ velocity field. %Black dots are fits to the \HI\ velocity field, while open dots are fits to the CO field.
The panels are (left) rotation velocity, (right) radial non-circular velocity, either `expansion' or `contraction'. PA is fixed at $296.1\degrees$ north of west and inclination is fixed at $i=59.2\degrees$. The rotation of the \HI\ rises slowly, and appears to be still rising to velocities $>200$ \kms\ at the last point measured. %We observe non-circular velocities at all radii within 25 kpc of the galaxy's center. %In contrast, the CO appears kinematically separate from the \HI, reaching a velocity of $200$ \kms\ within 10 kpc of the galaxy's center, and displays no significant non-circular motion.
 \label{fig:rotcur-UGC6168}}
\end{figure}

Figure \ref{fig:rotcur-UGC6168} presents the rotation curve of UGC 6168 derived from a tilted ring fit. The top left panel shows the rotation velocity as a function of radius. This rotation curve is derived with constant PA $=296.1\degrees$ north of west and $i=59.2\degrees$. UGC 6168's rotation increases slowly to $\sim200$ \kms\ over the 35 kpc which we can trace the gas, and appears to still be rising. Our \textsc{rotcur} model of the galaxy includes strong non-circular motions (top right panel) within 25 kpc of the galaxy's center. These non-circular velocities reach 40\% of the galaxy's rotation speed, significantly higher than the 9\% observed in the ``marginally unstable'' UGC 9037 by \citet{Hallenbeck2014}. It must be noted, however, that the sign of the non-circular velocities are degenerate with galaxy geometry, and these could be associated with either inward or outward flows. \textsc{rotcur} underestimates the uncertainties of the non-circular velocities. Fitting the galaxy's rotation curve with DiskFit---which more accurately estimates the uncertainties---indicate that models of radial or bar-like flows are of $1-2\sigma$ significance.

% We will return to this issue later. %The galaxy's position angle (bottom left), measured east of north, and inclination (bottom right) do not vary significantly over the disk. This is in agreement with observations that the inner disks of galaxies are very flat.

\begin{figure}
 \begin{center}
  \epsscale{1.0}
	\plotone{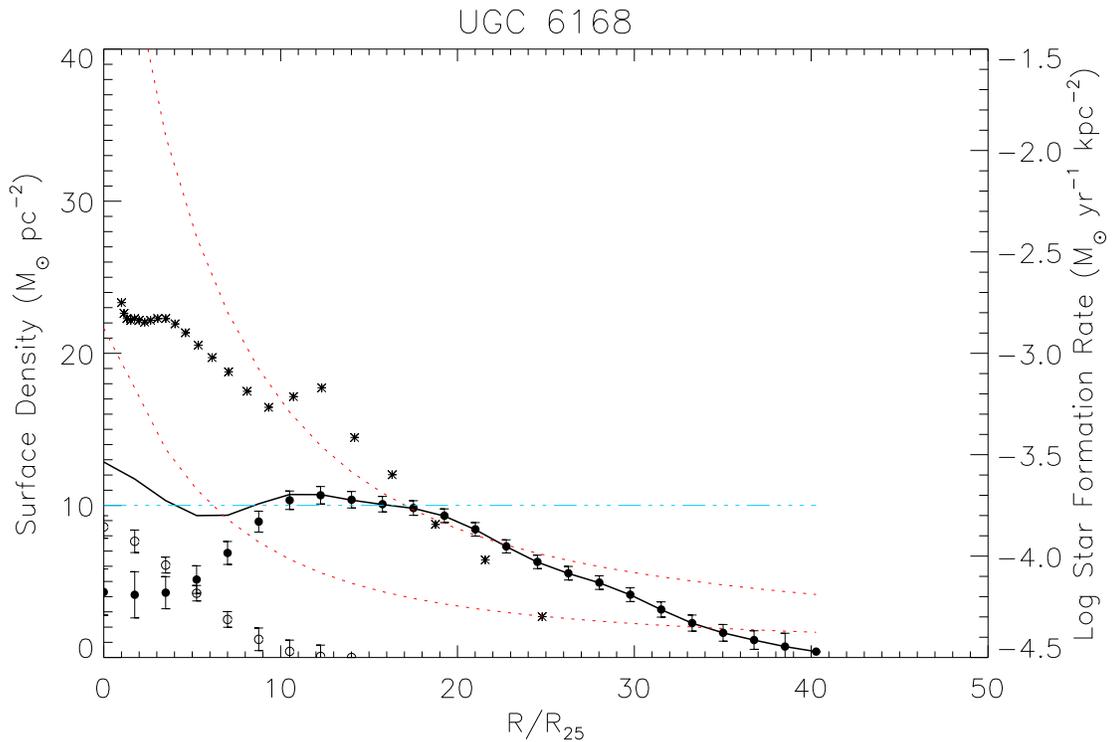}
 \end{center}
 \caption{Surface density profile of the \HI\ (closed circles) and \Htwo\ (open circles), as inferred from CO emission, of UGC 6168. The solid line is the total gas surface density. Stars are the H$\alpha$-derived star formation surface density (from \citealt{Huang2014}, Figure 15). Red dotted lines indicate where the gas surface density is 0.4 times (lower) and equal to (upper) the density at which the gas would be unstable according to its Toomre $Q$ parameter. Cyan dashed triple-dotted line is located at 10 \solarmassespersquareparsec, where \HI\ is observed to saturate in the local universe. At small galactocentric radii, there is little \HI, but the total gas surface density remains constant. Between 10 and 30 kpc, the galaxy is marginally unstable, and approaches instability near 20 kpc.
 \label{fig:density-UGC6168}}
\end{figure}

Figure \ref{fig:density-UGC6168} shows the deprojected surface density profiles of both the \HI\ (closed circles) and \Htwo\ (open circles). The solid black line is the total gas density. The stars depict the star formation surface density (from \citealt{Huang2014}, Figure 15). In the inner disk, the total gas density remains relatively constant at $\sim10$ \solarmassespersquareparsec\ (dashed-dotted cyan line), a surface density beyond which there is rarely \HI\ at solar metallicities (e.g., \citealt{Bigiel2008}); at solar abundance, gas above this threshold tends to be entirely molecular. This is the ``saturation line'' referred to in this and subsequent figures. The upper red dotted line indicates a gas surface density corresponding to an unstable ($Q<1$; upper line) thin gas disk. Generally, widespread disk instability is not observed in the local universe; any time gas surface densities exceed this value, it is likely a failure of the single phase thin gas disk model (for discussion of a more realistic \QN\ model for all three galaxies, see \S\ref{sec:multiphase}). However, at surface densities corresponding to $Q\lesssim2.5$ (lower red dotted line), star formation is observed to be enhanced (\citealt{Kennicutt1989}; \citealt{MartinKennicutt2001}; \citealt{Leroy2008}). We refer to $1<Q<2.5$ as the ``marginally unstable'' regime. For UGC 6168, the gas disk is predicted to be stable in the interior 7 kpc and at $r>18$ kpc, and is marginally unstable at intermediate radii. This marginal instability coincides with a SFR enhancement near 12-15 kpc. However, overall star formation is peaked at the center of the galaxy, where we predict the gas to be stable.
\newpage
\subsection{Overview of UGC 7899}

\begin{figure}
 \begin{center}
   \epsscale{0.75}
   \plotone{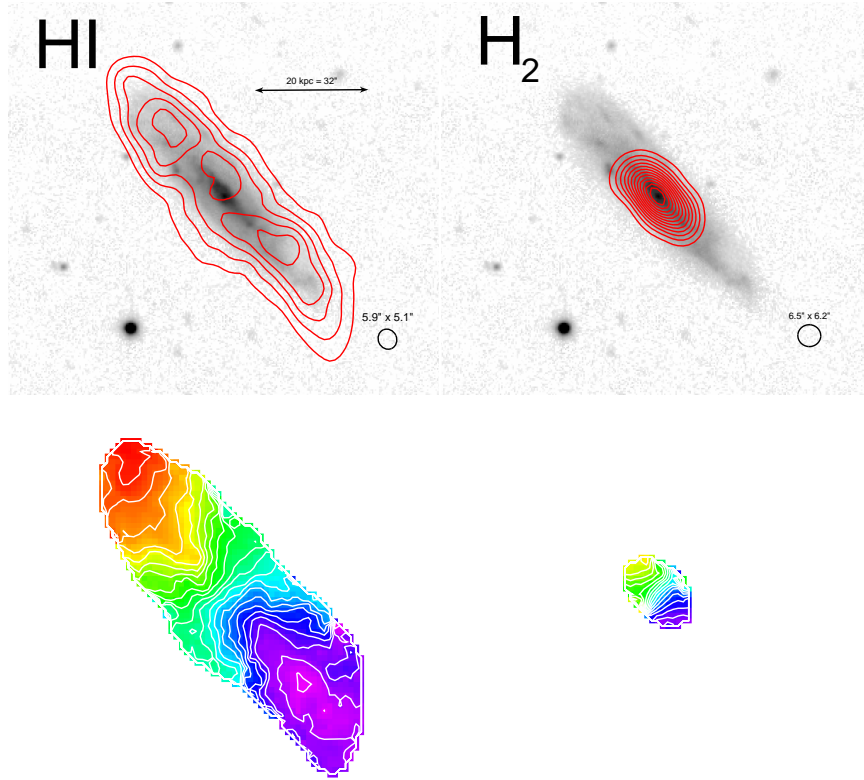}
 \end{center}
 \caption{Integrated flux maps and velocity fields of UGC 7899. Panels are identical to Figure \ref{fig:images-UGC6168}. The projected surface densities of CO in UGC 7899 are much higher than either UGC 6168 or NGC 5230, reaching 50\solarmassespersquareparsec ($\sim$25 \solarmassespersquareparsec\ deprojected). The \HI\ rotation shows some asymmetry: there are closed contours on the southern (approaching) half of the galaxy, indicative of a likely warp in the \HI\ disk at large galactocentric radii, which is absent in the northern (receding) half of the galaxy.\label{fig:images-UGC7899}}
\end{figure}

\noindent
Images, surface density contours, and velocity fields of UGC 7899 appear in Figure \ref{fig:images-UGC7899}; the panels are identical to Figure \ref{fig:images-UGC6168}. Like UGC 6168, UGC 7899 is an inclined spiral galaxy, with $i\sim72\degrees$. It also exhibits a color reversal, where the galaxy exhibits redder colors at larger radii than at intermediate radii \citep{Huang2014}. A close inspection also reveals that the optical galaxy is not perfectly symmetric: it tapers, with the optical emission at the northern edge of the galaxy more extended than at the southern edge.

The \HI\ and inferred \Htwo\ contours begin at projected 5 \solarmassespersquareparsec\ and increase by 5 \solarmassespersquareparsec\ at each additional contour. The \HI\ in the center of UGC 7899 is not depleted: there is no \HI\ ring. Instead, its surface density becomes approximately constant. There are, however, high \Htwo\ surface densities---much higher densities that are observed in either UGC 6168 or UGC 7899, reaching a projected column density of $\sim50$ \solarmassespersquareparsec\ (corresponding to roughly 25 \solarmassespersquareparsec\ when deprojected). The contours on the southern side of the galaxy suggest a slight warp in the \HI\ disk at large radii.

The bottom panels show the velocity fields of the \HI\ (left) and CO (right). The \HI\ velocity field appears slightly asymmetric, suggesting a warp in the outermost part of the disk. Like UGC 6168, the CO is only marginally resolved at the scale of the CARMA beam, and no rotation curve can be fit.

\begin{figure}
 \begin{center}
  \epsscale{1.0}
	\plotone{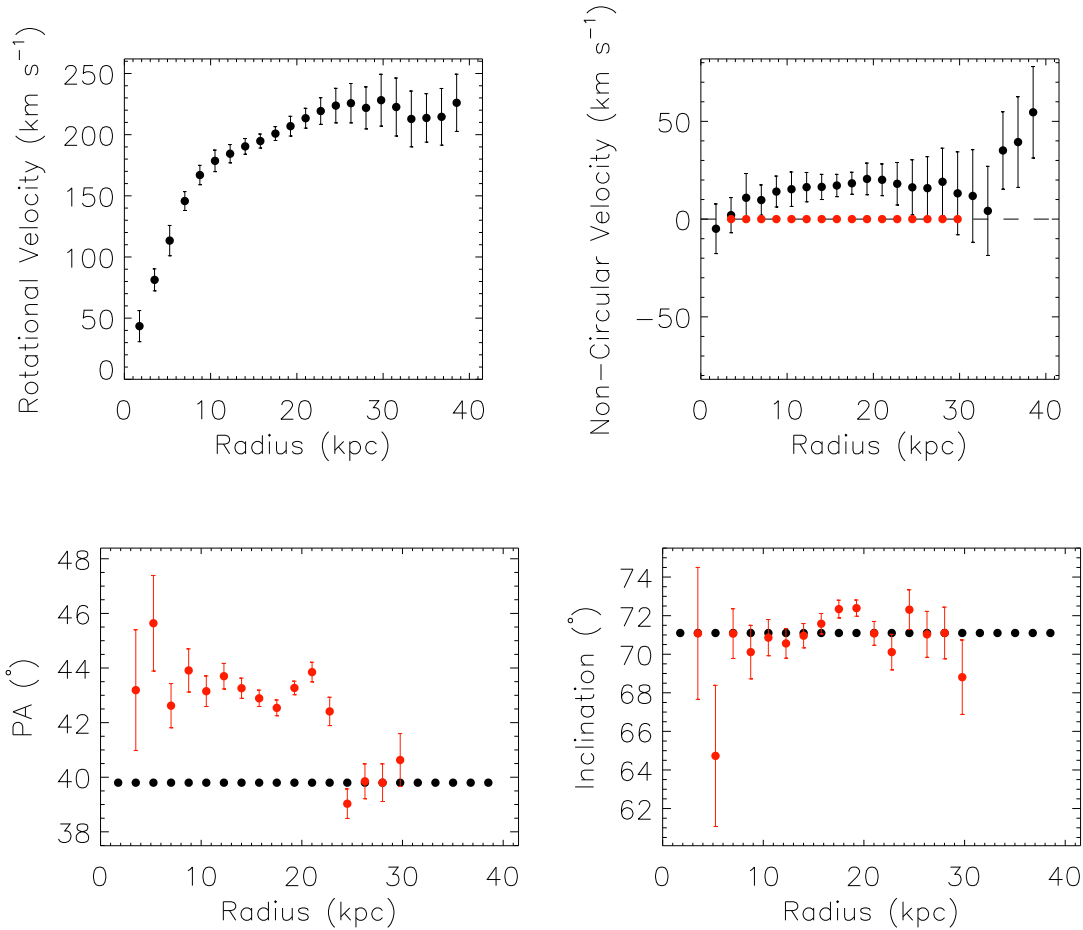}
 \end{center}
 \caption{Rotation curve of UGC 7899 derived by fitting a tilted ring model (black circles). Panels are identical to Figure \ref{fig:rotcur-UGC6168}. The rotation curve quickly rises to a maximum speed of 200-250 \kms\ at a radius of 10 kpc and remains flat. The non-circular velocities fit here prove to be only of marginal significance when re-examined with DiskFit. Instead, a model without non-circular velocities (red circles) is fit, which captures the warp with the abrupt change of the position angle at 25 kpc. %Some small, possibly significant non-circular motions are observed at radii of $\lesssim30$kpc.
 \label{fig:rotcur-UGC7899}}
\end{figure}

Figure \ref{fig:rotcur-UGC7899} presents the results of tilted ring fits to the UGC 7899's \HI\ velocity field (black circles). The best fit constant PA and $i$ are PA $=39.8\degrees$ north of west and $i=71.1\degrees$. The rotation curve rises linearly to $\sim180$ \kms\ then changes to a shallower slope, reaching a maximum of $\sim230$ \kms. The fit becomes uncertain at large radii because of the north-south mismatch mentioned above: we find a declining rotation curve in the southern approaching half of the galaxy, but not in the northern receding half. DiskFit suggests that these observed non-circular flows are of marginal ($1-2\sigma$) significance, and only in the outer regions. Because geometry and non-circular velocities are degenerate, we fit a second model to the map to capture the effect of the galaxy's warp (red circles). This model holds fixed both the rotation curve and a radial flow of 0 \kms. Here we observe a relatively constant position angle, which changes abruptly by $4\degrees$ at a galactocentric radius of 25 kpc.

%There are non-circular velocities over the entire galaxy disk, reaching a maximum of $\sim10$\% of the galaxy's rotation speed, similar to those seen in HIghMass galaxy UGC 9037. %The inclination and position angles are generally stable, only varying by 5\degrees over the entire galaxy disk, suggesting a physical very flat disk. The only trend in the fit is a linear change in inclination between 10 and 20 kpc. However, forcing the galaxy to a constant inclination angle and position angle over that range has no significant effect on the rotation and non-circular velocities.

\begin{figure}
 \begin{center}
  \epsscale{1.0}
	\plotone{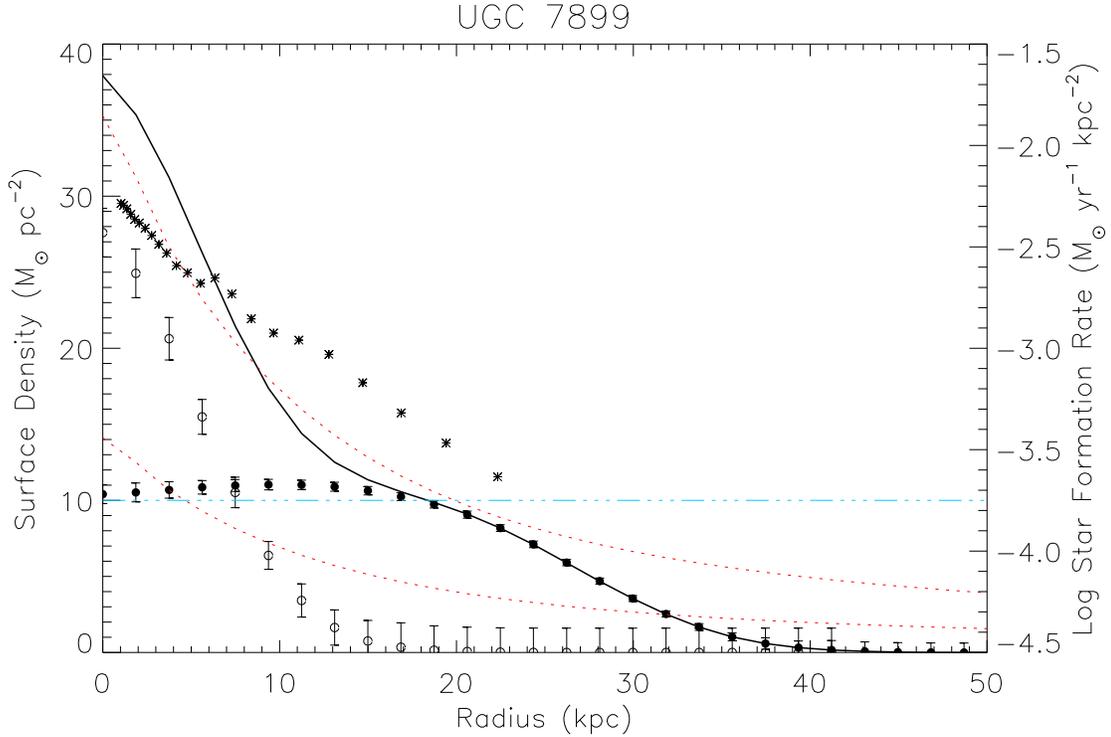}
 \end{center}
 \caption{Surface density profile of the \HI\ and \Htwo\ in UGC 7899. Symbols are identical to Figure \ref{fig:density-UGC6168}. Unlike UGC 6168 and NGC 5230, there is no hole in the \HI. The \Htwo\ surface densities reach 28 \solarmassespersquareparsec, higher than either of the other two galaxies in this work by 50\%. The gas appears marginally unstable according to Toomre $Q$ at all radii inside of 30 kpc, and becomes unstable in the inner 10 kpc, where the \Htwo\ dominates the gas surface density.
 \label{fig:density-UGC7899}}
\end{figure}

Figure \ref{fig:density-UGC7899} presents de-projected surface densities of \HI\ and \Htwo\ as a function of radius. Also included are our $Q<1$ and $Q<2.5$ stability curves and a $\Sigma_{HI}=10$ \solarmassespersquareparsec\ saturation line. Unlike UGC 6168, there is no \HI\ hole in the center of the galaxy. Instead, the \HI\ saturates at 10 \solarmassespersquareparsec, while $\Sigma_{\text{H}_2}$ reaches nearly 30 \solarmassespersquareparsec, 50\% higher than either of the other two galaxies in this work. We predict that the disk is marginally unstable for $r<30$ kpc, with the disk at or near instability for $r<20$ kpc. This is in agreement with the observation that the most active star formation in UGC 7899 is centrally located.%, and our observations that the gas is experiencing non-circular motions at nearly all radii.% It cannot explain, however, why the galaxy is bluest at approximately 20 kpc---the stability criteria suggest that the galaxy should be very blue within the entire inner 20 kpc of the galaxy.

\subsection{Overview of NGC 5230 (UGC 8573)}

\begin{figure}
 \begin{center}
   \epsscale{0.75}
   \plotone{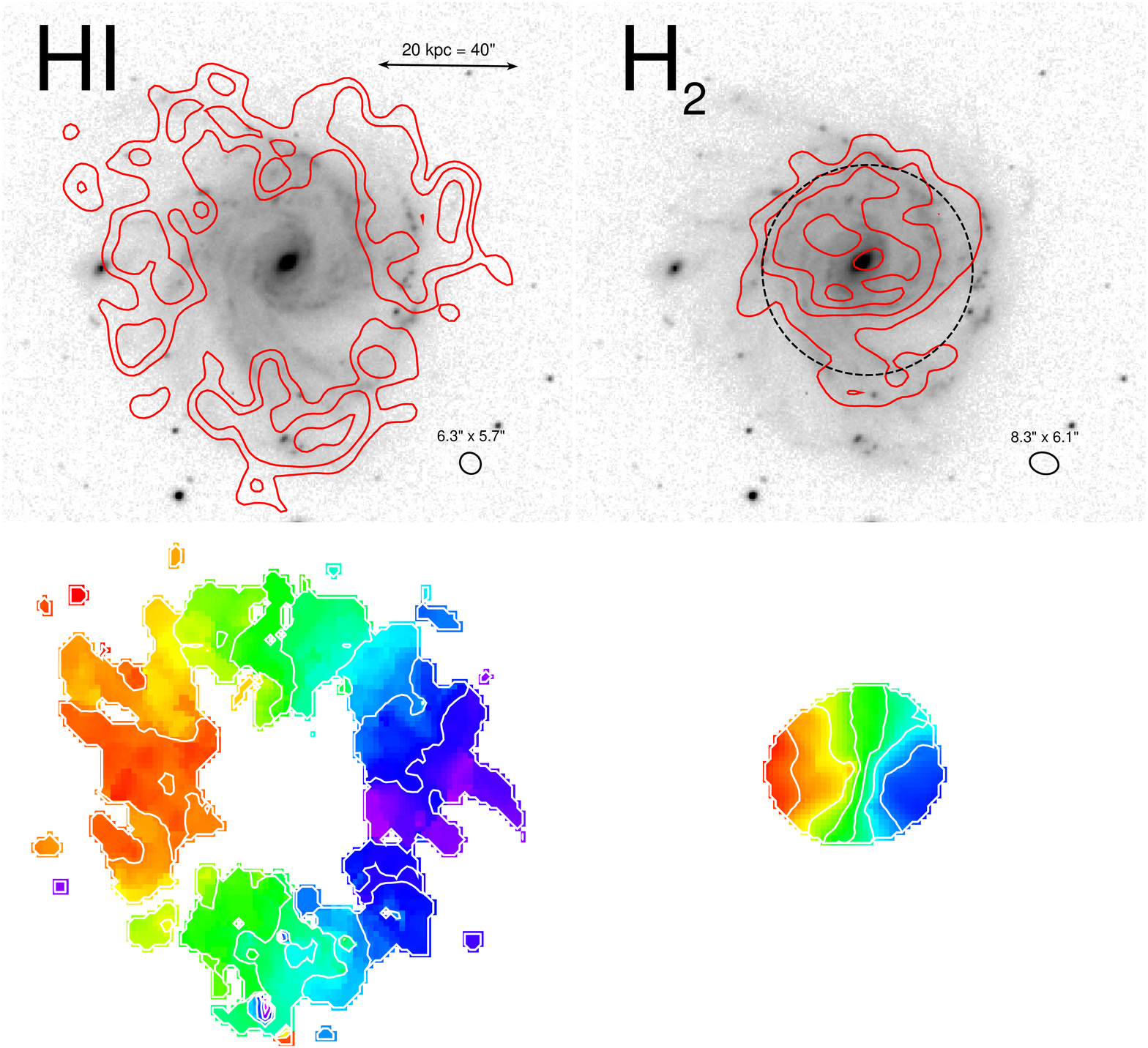}
 \end{center}
 \caption{Integrated flux maps and velocity fields of NGC 5230. Panels are identical to Figure \ref{fig:images-UGC6168}. The inclination of NGC 5230 is lower than for the other two galaxies ($i\sim40\degrees$), allowing a clear view of the \HI\ ring. CO emission reaches beyond the 50\% response radius of CARMA ($\sim30^{\prime\prime}$; dashed line), but not beyond the 20\% response radius. The CO emission is resolved enough by CARMA to produce a rotation curve.%, and its twisted contours show significant non-circular velocities.
\label{fig:images-UGC8573}}
\end{figure}

\noindent Figure \ref{fig:images-UGC8573} presents the \HI\ and CO moment maps of NGC 5230. %Unlike the other two galaxies discussed in this work, NGC 5230 was observed using the GMRT, and followed a slightly different AIPS pipeline to produce the moment maps. Its analysis was otherwise identical.
 NGC 5230 is a large spiral galaxy with a low inclination ($i\sim20\degrees$). Its southern arm appears less tightly wound than the other two in the north. Like UGC 6168, there is a large central hole where the \HI\ has essentially all been converted into \Htwo. However, the \HI\ hole in NGC 5230 is larger, with a radius of 10-15 kpc and has no \HI\ down to our detection limits of 1.63 \solarmassespersquareparsec. All three spiral arms show enhanced densities of \Htwo, out to CARMA's half power response radius (dashed line). We thus cannot know if all of the CO has been mapped. %Due to the The CO emission may be extendedBecause of how extended the CO emission is%, it is unclear whether we have recovered  %However, a second, poorer quality pointing suggests that there is essentially no emission beyond the 20\% response radius.
Regardless, we have adequately mapped the CO emission of the inner 30'' ($\sim15$ kpc). %, and our inferred \Htwo\ mass can be considered at worst a lower limit.
%In addition, a rotation curve can be independently fit to the CO emission. %The velocity fields of both the \HI\ and CO velocity contours show some deviation from perfect circular rotation, more notably in the CO, where the isovelocity contours twist significantly.

Because of the low inclination of NGC 5230 ($i < 40^\circ$), fitting a rotation curve and deriving surface densities from \HI\ velocity fields via tilted ring fits \citep{rotcur} is highly unreliable. We thus remove it from our later surface density, rotation curve, and dark matter fitting except as noted.

\subsection{Multi-Phase Stability Parameter}
\label{sec:multiphase}
\begin{figure}
 \begin{center}
  \epsscale{1.0}
	\plotone{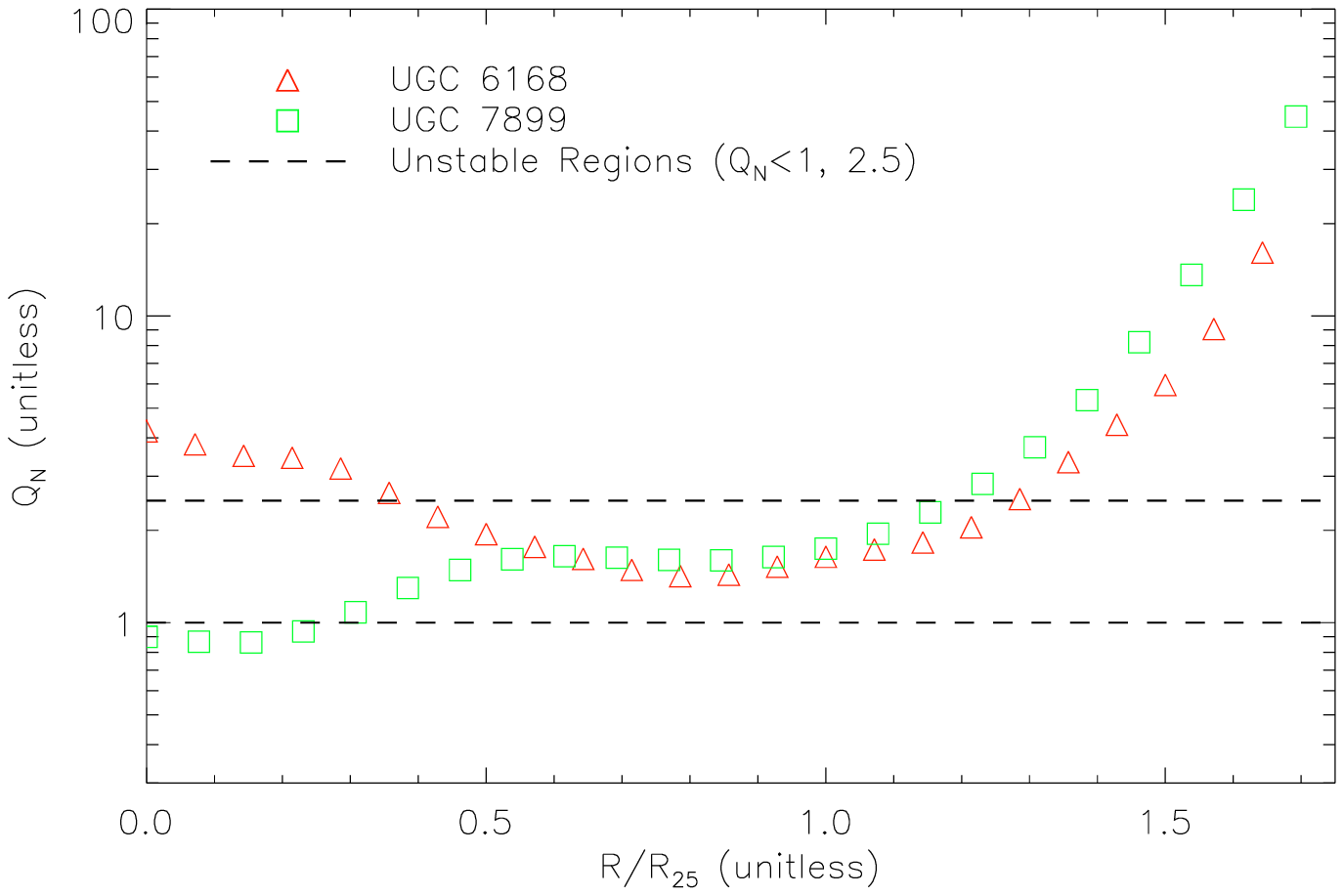}
 \end{center}
 \caption{Multi-phase \QN\ from \citet{RomeoFalstad2013} for the three HIghMass galaxies in this paper as a function of galactic radius. Radii are normalized to $r$-band $R_{25}$. Red open triangles are UGC 6168 and green open boxes are UGC 7899. %, and blue crosses are NGC 5230.
 The black dashed line separates the unstable ($\QN<1$) and marginally stable ($\QN\lesssim2.5$) from the stable ($\QN>2.5$) regions.\label{fig:Q}}
\end{figure}

\noindent
Figure \ref{fig:Q} presents the multi-phase \QN\ of \citet{RomeoFalstad2013} for both UGC 6168 (red triangles) and UGC 7899 (green boxes). The radii have been normalized by $r$-band $R_{25}$, the radius at which the isophotes reach 25 mag.arcsec.\per{2}. The dashed line indicates a value of $\QN=1$. For values of $\QN<1$, the disk is unstable to perturbations, while for $\QN>1$, the disk is stable.

Both galaxies show similar trends: they are predicted to be stable beyond the optical disk ($R>R_{25}$), and are marginally unstable ($\QN\lesssim2.5$) in the region $0.5<R/R_{25}<1.0$. In the inner disk ($R<0.5 R_{25}$), UGC 6168 becomes stable again, but the high surface densities of \Htwo\ cause UGC 7899 to become unstable. Neither disk is predicted to be unstable to ring-like perturbations.

These results are not greatly different from the one-phase Toomre Q results discussed in the previous sections. The only significant change is that for UGC 6168, the multi-phase model predicts a marginally unstable ($\QN\lesssim2.5$) disk at intermediate radii, while the single-phase predicts a fully unstable disk. %We note once again the surprising and likely unrealistic result that NGC 5230 is predicted to be unstable ($\QN<1$) over virtually all radii.

\subsection{HI Radii}

\begin{figure}
 \begin{center}
  \epsscale{1.0}
	\plotone{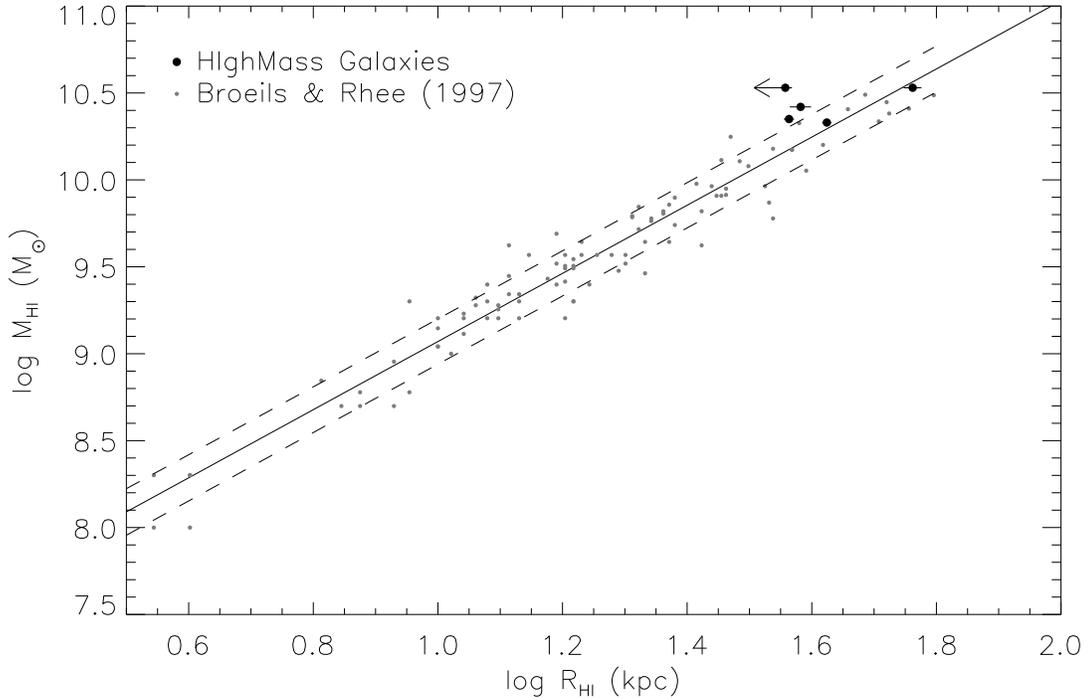}
 \end{center}
 \caption{\HI\ mass as a function of radius for the spiral and irregular galaxies of \citealt{BroeilsRhee1997} (small gray dots), along with best-fit line; dashed line is the $1\sigma$ of the line. The larger black dots are HIghMass galaxies. The three HIghMass galaxies presented in this paper along with UGC 9037 lie to the left of the best-fit line (i.e. they have smaller $R_{HI}$ than predicted), while only the low surface brightness galaxy UGC 12506 is within error of the line. For some of the HIghMass galaxies, the uncertainties are smaller than the plotted points.\label{fig:broeils-rhee}}
\end{figure}

\noindent
For both typical spiral galaxies as well as more massive spiral galaxies, there exists a tight linear correlation between the total \HI\ gas mass and the radius at which the deprojected surface density reaches 1 \solarmassespersquareparsec\ (\citealt{BroeilsRhee1997}; \citealt{Bluedisks}; \citealt{HISizeMass}). This relationship can be expressed as (from Equation 13 in \citealt{BroeilsRhee1997}):
\begin{equation}
 \label{eqn:hiradii}
 \log \frac{\hat{R}_\text{\HI}}{\text{kpc}} = 0.51 \log \frac{M_\text{\HI}}{M_\odot} - 3.63.
\end{equation}
\noindent
%\citet{Hallenbeck2014} showed that this relationship also holds true for two of the galaxies in the HIghMass sample, despite their very different gas properties. One, UGC 9037, has significant non-circular velocities like the three galaxies presented in this work, while the other, UGC 12506, has very low surface densities and appears to be stable.
Figure \ref{fig:broeils-rhee} presents this line, along with the \citealt{BroeilsRhee1997} sample (gray dots). The HIghMass galaxies are plotted in black. This relationship also holds true for the galaxies discussed by \citet{Hallenbeck2014}---UGC 9037 and UGC 12506. The \HI\ disks of UGC 6168, UGC 7899, and NGC 5230 lie to the left of the line, and so are more compact than expected. For NGC 5230, we have plotted assuming that $i= 0^\circ$, which yields the largest value of $R_\text{HI}$. For the three galaxies this difference is near the edge of significance ($1.5\sigma-2.9\sigma$), but taken together, the difference is reasonably significant ($3.6\sigma$). Theoretically, the baryonic disk scale radius should increase with $\lambda$ and the rotational velocity of the disk (e.g. \citealt{MoMaoWhite1998}, \citealt{Hernandez2007}, \citealt{Berta2008}). This could indicate that these galaxies have somehow had their \HI\ compressed, or that instabilities in the disk are allowing the normally extended, low surface density gas to flow inward. Table 2 presents the \HI\ radii of all five galaxies, along with their expected \HI\ radii.

\begin{deluxetable}{ccc}
\tablecolumns{3}
\tablewidth{0pt}
\tabletypesize{\scriptsize}
\tablecaption{Observed and Expected \HI\ Radii}
\tablehead{
 \colhead{Galaxy} & \colhead{$R_\text{\HI}$} & \colhead{$\hat{R}_\text{\HI}$}\\
                  & (kpc)                      & (kpc)\\
                  & (1)                        & (2)}
\startdata
UGC 6168          & 36.6 ($\pm0.7$)            & 45.0 ($\pm5.2$)\\
UGC 7899          & 38.2 ($\pm1.5$)            & 48.9 ($\pm5.7$)\\
NGC 5230          & 36.1 ($\pm0.9$)$^*$        & 55.6 ($\pm6.6$)\\
\hline
UGC 9037          & 42.1 ($\pm0.7$)            & 43.9 ($\pm3.5$)\\
UGC 12506         & 57.8 ($\pm1.9$)            & 55.6 ($\pm4.5$)
\enddata
\tablecomments{Observed and expected \HI\ radii of the three galaxies in this work and in \citet{Hallenbeck2014}. Column 1: observed radius where the \HI\ reaches a deprojected surface density of 1 \solarmassespersquareparsec, based on fitting the surface density profiles; Column 2: predicted \HI\ radius, based on Equation \ref{eqn:hiradii} and \citet{BroeilsRhee1997}.\\$^*$This is an upper limit assuming the inclination of NGC 5230 is face-on ($i = 0^\circ$) \label{tab:hiradii}}
\end{deluxetable}

%UGC 9037 also lies to the left of the line, though not significantly so. UGC 12506, a low surface brightness galaxy with an extended \HI\ disk, agrees with the best-fit line.

%All three galaxies have radii which are $\sim10$ kpc less than  predicted, although the difference is near the edge of significance ($1.6\sigma-2.4\sigma$) in each case. As these three galaxies were chosen from the HIghMass because of their unusually blue colors, we may suspect that their \HI\ properties may be unusual for the HIghMass sample, and consider them together. Under this assumption, together they have significantly smaller \HI\ radii than other galaxies of similar \HI\ mass ($3.4\sigma$).

\subsection{Star Formation Efficiencies}
\noindent
It has already been observed that the HIghMass galaxies have moderate to high values of SFE(\HI)$\equiv\text{SFR}/{M_\text{\HI}}$ relative to the ALFALFA sample as a whole \citep{Huang2012a}. In addition, our CARMA observations show that their star formation rates are also typical for their \Htwo\ masses: the \Htwo\ depletion timescales timescales ($\tau_{\text{H}_2}\equiv M_{\text{H}_2+\text{He}}/\text{SFR}$) are $0.3$, $0.4$, and $1.6\times10^{9}$ yrs for UGC 6168, UGC 7899, and NGC 5230, respectively. These values are below the timescale observed by the THINGS survey of $1.3-3.6\times10^{9}$ yrs \citep{Leroy2008}, as well as the $2.35\times10^9$ yrs found by \citet{Bigiel2011}. But, they overlap the predicted range found by the COLD GASS survey for galaxies of similar stellar masses (\citealt{COLD-GASS}; \citealt{COLD-GASS-ii}; $0.8-1.25\times10^{9}$ yrs).

However, the globally averaged SFE of \HI\ and \Htwo\ are just part of the story: %the averaged SFE(\HI) that an integrated profile probes indicative of low SFE(\HI) locally, or is it globally inefficient?
the resolved SFE of our sample is likely to shed more light.
%However, it is unclear what bottleneck is causing this: can \HI\ not convert to \Htwo\ efficiently, or is it in the formation of stars from clouds if \Htwo?
The THINGS survey found several ways to parametrize the efficiency of the total gas (\HI, \Htwo, and He) as a function of galactic radius, normalized by $R_{25}$ (see Equation 21 of \citealt{Leroy2008}). If we just consider the spiral galaxies in THINGS:
\begin{equation}
 \label{eqn:SFE}
 \text{SFE(gas)} = \left\{
  \begin{array}{lr}
	 5.9\times10^{-10} & R < 0.4R_{25}\\
	 3.0\times10^{-10}\exp\left(-\frac{R}{0.25R_{25}}\right) & R > 0.4R_{25}
	\end{array}
	\quad\text{yr}^{-1}\right.
\end{equation}
\noindent
This form represents a constant value of $SFE(\text{\Htwo})$, and thus a constant $SFE(\text{gas})$ in the region of the interstellar medium which is \Htwo\ dominated ($\Sigma_\text{\Htwo}>\Sigma_\text{\HI}$; $R\lesssim0.4R_{25}$). Outside of this region, the disk is \HI\ dominated and the conversion of gas to stars becomes inefficient, declining exponentially with radius.

\begin{figure}
 \begin{center}
  \epsscale{1.0}
	\plotone{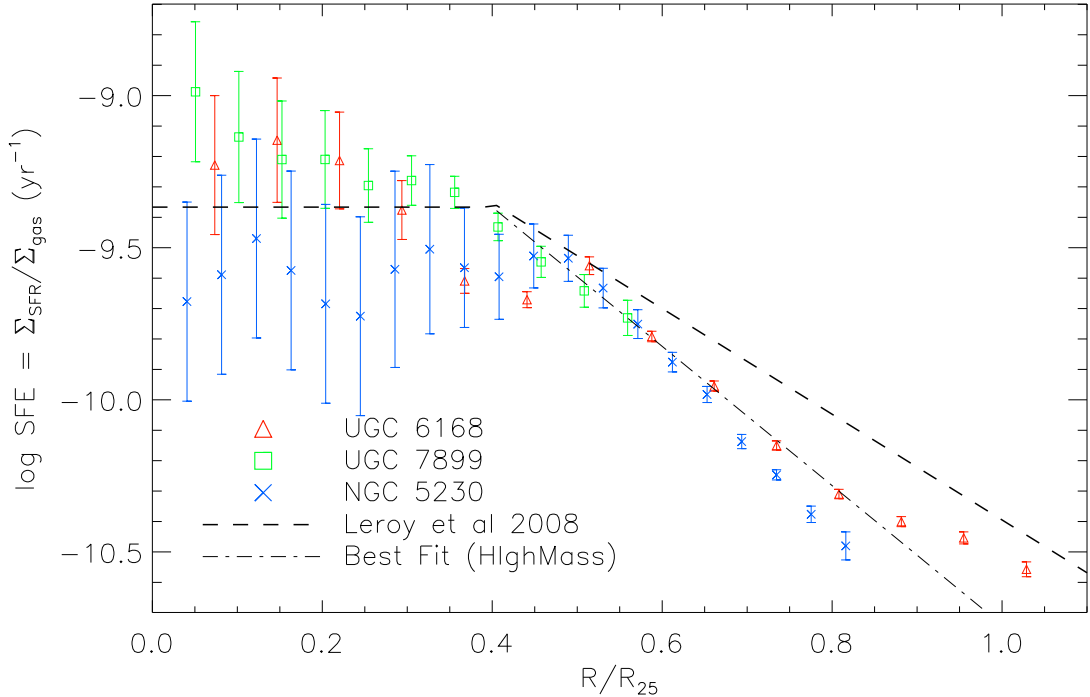}
 \end{center}
 \caption{Total gas (\HI, \Htwo, and He) star formation efficiency as a function of normalized galactic radius. Red open triangles are UGC 6168, green open boxes are UGC 7899, and Blue crosses are NGC 5230. The black dashed line represents a best-fit model for massive spirals found by THINGS \citep{Leroy2008}. The kink is near the transition from an \HI-dominated to an \Htwo\ dominated ISM; in the \HI-dominated region, the SFE falls off exponentially with a scale length of $0.25R_{25}$. In the \Htwo-dominated central regions of each galaxy, the HIghMass galaxies follow the same trend as the massive THINGS spirals, but in the \HI-dominated outskirts, the HIghMass galaxies are all significantly less efficient at forming stars. The narrower dash-dotted line is a best fit to the HIghMass galaxies alone, which have a shorter exponential scale length of $0.190(\pm0.001)R_{25}$. \label{fig:SFE}}
\end{figure}

Figure \ref{fig:SFE} presents this fit, along with SFE(gas) for the three HIghMass galaxies in this work. Red triangles, green boxes, and blue crosses represent data for UGC 6168, UGCC 7899, and NGC 5230, respectively. Here, we include NGC 5230 because SFE(gas) depends on the ratio of two surface densities, and so is independent of the assumed inclination. The dashed line is the parameterization in Equation \ref{eqn:SFE}. The HIghMass galaxies in this work all exhibit the same overall trend of the THINGS parameterization: there is a core with nearly constant SFE(gas), and then at some radius, SFE begins to drop. At small radii, where the \Htwo\ dominates the ISM, we see essentially no difference between the galaxies in our sample and the average spiral galaxy in the THINGS survey. In the outer disk, where the \HI\ dominates the ISM, the SFE of our sample is significantly lower than for THINGS. The dashed-dotted line is a best-fit to the HIghMass data, assuming an exponential decrease in SFE(gas) beyond $R>0.4R_{25}$. This has a significantly shorter scale length of $0.190(\pm0.001)R_{25}$.

Given that the global SFE(\HI) for these galaxies is typical, it is somewhat surprising to find that the resolved SFE(gas) is low where the ISM is primarily \HI. This may be a consequence of the gas being unusually compact in these galaxies. The \Halpha\ emission has been traced to only approximately $R/R_{25}<1.0$. Of the total \HI\ mass, 30\% resides beyond this radius for UGC 6168 but only 3\% for NGC 5230. For UGC 7899, this explanation is less conclusive: the star formation is only traced to $R/R_{25}<0.6$ with 70\% of the total \HI\ mass residing beyond that radius.

A related possibility is that the relatively lower mass spiral sample of THINGS is a poor comparison sample. $R_{25}$ for the THINGS galaxies varies between 10 and 20 kpc while the three galaxies in this work have radii of $24-36$ kpc (the HIghMass sample as a whole has an average $R_{25}$ of 20 kpc). The THINGS-derived SFE(gas) relationship has a factor of 2-3 scatter, and in general the galaxies with larger $R_{25}$ in the THINGS sample tend to fall below the relationship \citep{Leroy2008}. This may hint at a possible decrease in SFE scale radius at large galaxy radii.

\subsection{Dark Matter Properties}
\noindent%\label{DMprops}
%We use each galaxy's rotation curve along with mass models of the baryonic components to determine the dark matter content of each galaxy.
We model each galaxy with four components: an \HI\ disk, an \Htwo\ disk, a stellar disk, and dark matter. We use the \HI\ and \Htwo\ surface densities derived in \S\ref{sec:data} and \S\ref{sec:COdata} and treat each as a thin disk. The stellar component is also modeled as a thin disk, using the stellar masses described in \S\ref{sec:stellarmass}. The contribution of the dark matter can be determined via:
%We fit both Navarro-Frenk-White (NFW) and pseudo-isothermal (ISO) dark matter models to UGC 6168 and UGC 7899. Our models consist of four components: a stellar disk, an \HI\ disk, an \Htwo\ disk, and the dark matter profiles (the \HI\ and \Htwo\ disks have their masses corrected for the presence of Helium).
\begin{equation}
	V^2_\text{obs} = V_\text{HI}^2 + V^2_{\text{H}_2} + V^2_* + V^2_{DM}
\end{equation}
where $V_\text{obs}$ is the rotation curve, and the other velocities are the contributions from each phase. Both a Navarro-Frenk-White (NFW) and pseudo-isothermal profile (ISO) are separately fit to the remaining $V_{DM}$ component.
%As is typical, the stellar mass is much more centrally concentrated than either gas phase, and so dominates the fit in the inner $5-10$ kpc.

Using the methods of \citet{Querejeta2014} predict mass to light ratios ($\Upsilon_{3.6}$) which are too high: the resulting stellar mass surface densities yield $V^2_*>V^2_\text{obs}$ or unphysical parameters for dark matter fits ($c<0$, for example). The $\Upsilon_{3.6}$ distribution has a spread of roughly 0.1 dex, which is insufficient to explain the difference. Instead, this is likely because we have not accounted for the effect of dust or stars of intermediate age contaminating the Spitzer $3.6\mu m$ and $4.5\mu m$ bands, which together can account for 20-60\% of the total infrared emission \citep{Meidt2012}. If we allow $\Upsilon_{3.6}$ to vary from its nominal value, best fits for both galaxies yield similarly inappropriate $\Upsilon_{3.6} \approx 0$. Instead, we set the average value of $\Upsilon_{3.6}$ such that the Spitzer-derived mass equals the mass from SED-fitting. In such cases, we get good fits for both UGC 6168 and UGC 7899. This requires values of $\Upsilon_{3.6}$ a factor of 2-3 lower than the nominal values derived by \citet{Querejeta2014}.
%
%So, as is standard, we allow the contribution of the stellar mass to the disk to vary by adding a dimensionless term: $V^2_* \rightarrow m_D V^2_*$ in all equations.

\placetable{tab:darkmatter}

\begin{deluxetable*}{c|cc|ccc|ccc|c}
\tablecolumns{10}
\tablewidth{0pt}
\tabletypesize{\scriptsize}
\tablecaption{Dark Matter Fits}
\tablehead{
 Galaxy           &                               &                                          &                   & NFW Fit       &               &                               & ISO Fit       &                 & \colhead{Halo Spin}\\
                  & $\left<\Upsilon_{3.6}\right>$ & $\left<\Upsilon_{3.6}^\text{SED}\right>$ & $c$               & $R_{200}$     & $\chi^2_\nu$  & $\rho_{C}$                    & $R_{C}$       &  $\chi^2_\nu$   & $\lambda$ \\
                  & ($M_\odot/L_\odot$)           & ($M_\odot/L_\odot$)                      &                   & (kpc)         &               & ($10^{-3} M_\odot$ pc\per{2}) & (kpc)         &                 & \\
                  & \colhead{(1)}                 & \colhead{(2)}                            & \colhead{(3)}     & \colhead{(4)} & \colhead{(5)} & \colhead{(6)}                 & \colhead{(7)} & \colhead{(8)}   & \colhead{(9)}}
\startdata
UGC 6168          & 0.45                          & 0.27                                     & $1.47\pm0.51$     & $203\pm36$    & $1.03$        & $9.2\pm1.6$                   & $10.4\pm1.3$  & $0.41$          & $0.09$\\
UGC 7899          & 0.43                          & 0.16                                     & $1.75\pm0.29$     & $224\pm23$    & $0.10$        & $13.60\pm0.52$                & $9.00\pm0.28$ & $0.02$          & $0.08$\\
\enddata
\tablecomments{Best fit results of dark matter halo models to UGC 6168 and UGC 7899, using either a Navarro Frenk White (NFW) or pseudo-isothermal (ISO) dark matter halo model. Column 1: Nominal average mass-to-light ratio of each galaxy in the 3.6\micron\ band; Column 1: adopted mass-to-light ratio from setting total stellar mass from Spitzer observations equal to that calculated via SED fitting; Columns 3 and 4: concentration index and characteristic halo length scale; Column 5: reduced $\chi^2$ of NFW fit; Column 6-7: halo core density and length-scale; Column 8: reduced $\chi^2$ of ISO fit; Column 9: modified halo spin parameter, based on ISO fit.}
\label{tab:darkmatter}
\end{deluxetable*}

%Table \ref{tab:darkmatter} presents the results of fitting for each of the galaxies presented here. In each case, we find $\Upsilon_{3.6}$ is consistent with zero, that is, we can only fit dark matter profiles when the stellar disk gives a negligible contribution to the overall kinematics of the galaxy. Additionally, despite their large uncertainties, these values disagree with those predicted by \citet{Querejeta2014}. This was not the case for either of the previously studied HIghMass galaxies \citep{Hallenbeck2014}. The discrepancy may be in how the stellar mass surface densities are derived: we are using stellar masses derived from Spitzer IRAC observations, while \citet{Hallenbeck2014} used SED fitting to SDSS photometry. For UGC 6168, UGC 7899, and NGC 5230, the IRAC stellar masses are larger by a factor of $2-3$ compared with the SED fit masses.

\begin{figure}
 \begin{center}
  \epsscale{1.0}
	\plotone{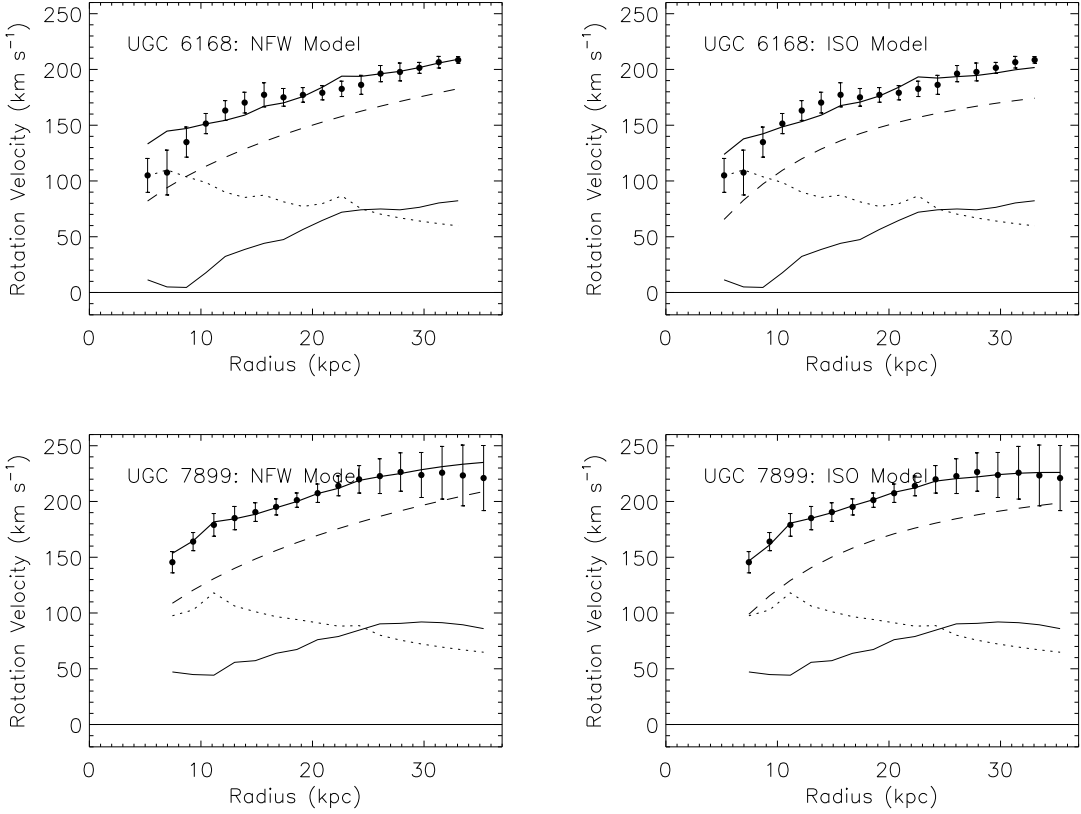}
 \end{center}
 \caption{Rotation curves (black dots) and best-fit mass models (thick black lines) for UGC 6168 and UGC 7899. Long dashed lines are dark matter halo fits using an NFW (left) or pseudo-isothermal (right) profiles. The thin black line is the total gas (\HI, \Htwo, and He) contribution to the rotation curve, while the short dashed line is the contribution from the stellar disk, after fitting $\Upsilon_{3.6}$. The dark matter is the dominant contributor to the rotation curve for every model. Fits for UGC 6168 and UGC 7899 are excellent for both halo profiles ($\chi^2_\nu \sim 0.5$).\label{fig:massmodels}}
\end{figure}

In Figure \ref{fig:massmodels}, we present both the dark matter fits to each galaxy's rotation curve, using the NFW (left) and pseudoisothermal (right) halo profiles. Black dots with uncertainties are the observed rotation curve. The thin solid black line is the total gas (\HI$+$\Htwo$+$He) contribution, the short dashed line is the stellar mass contribution, and the long dashed line is the halo fit. The thick solid line is the summed contribution from each of the mass terms. For each galaxy, the dark matter is the dominant contribution to the rotation curve in the best fit model at all radii. For UGC 6168 and UGC 7899, both the ISO and NFW profiles fit well ($\chi^2_\nu \sim 1$). Numerical results of the dark matter fits can be found in Table 3. %For NGC 5230, all fits are poor ($\chi^2_\nu \sim 5$).% The lack of a good fit in NGC 5230 is likely related to the kink in the gas' contribution to the rotation curve at 12 kpc, where the disk transitions from \HI\ to \Htwo\ dominated. This kink persists regardless of inclination assumed for NGC 5230.

\begin{figure}
\begin{center}
 \epsscale{1.0}
 \plotone{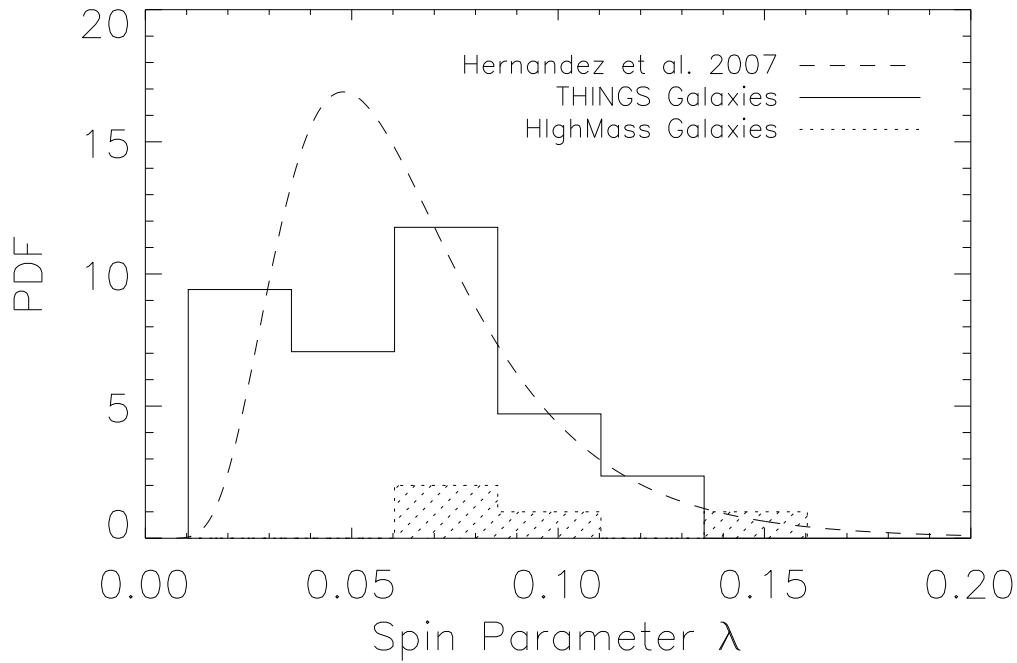}
\end{center}
\caption{Dark matter halo spin parameters of three samples. The solid histogram is the THINGS sample, calculated from both local optical and gas properties according to the method of \citet{Hallenbeck2014}, with area normalized. The dashed line is the best-fit probability density function calculated by \citet{Hernandez2007}. That work uses global optical properties alone. The filled dotted histogram is from the sample in this work combined with \citet{Hallenbeck2014}. There are two clear subsamples of HIghMass galaxies: UGC 12506 has a high spin parameter (0.15), while UGC 9037, 6168, and 7899 have intermediate values. \label{fig:spins}}
\end{figure}

Figure \ref{fig:spins} plots a histogram of dark matter halo spin parameters of three different samples. The solid histogram is the THINGS sample calculated as described in \citet{Hallenbeck2014}, and is normalized to unit area. For comparison is the volume-limited sample from the SDSS taken by \citet{Hernandez2007} and calculated from the global optical properties alone. The two distributions (and thus calculation methods) are roughly the same, especially accounting for the differences in sample size (19 THINGS galaxies versus 11597 SDSS-selected galaxies). The remaining filled dotted bars are non-normalized counts of the galaxies from this work combined with \citet{Hallenbeck2014}. %Compared against either sample, NGC 5230 is a clear low-spin outlier ($\lambda = 0.02$), although this low value may be due to the uncertainty in the inclination of the galaxy.
UGC 12506 clearly has a high spin ($\lambda=0.15$). UGC 6168, 7899, and 9037 all have values which are somewhat high, but not far into the tail ($\lambda \approx 0.09$).

%Our halo spin parameters are calculated using the maximum velocity the dark matter halo reaches within the galaxy, not its maximum value (see \S 3.14 of \citealt{Hallenbeck2014}). For NGC 5230, like UGC 9037 and UGC 12506, the rotation curve is very flat, and so this assumption has negligible impact on our calculated spin parameters. However, for UGC 6168 and 7899, the rotation curves are consistent with rising at the last measured point (see Figures \ref{fig:rotcur-UGC6168} and \ref{fig:rotcur-UGC7899}). If we instead use the fitted maximum circular velocity of the halo, both values of $\lambda$ increase by $\sim10\%$ to $0.10$. Even with this change, they remain distinctly lower than the $\lambda=0.15$ of UGC 12506.

%Dark matter halo spin parameters of three samples. The solid histogram is the THINGS sample, calculated from both local optical and gas properties according to the method of \citet{Hallenbeck2014}, with area normalized. The dashed line is the best-fit probability density function calculated by \citet{Hernandez2007}. That work uses global optical properties alone. The filled dotted histogram is from the sample in this work combined with \citet{Hallenbeck2014}. There are three clear subsamples of HIghMass galaxies: NGC 5230 has a low spin parameter ($\lambda=0.019$), UGC 12506 has a high spin parameter (0.153), while UGC 9037, 6168, and 7899 have intermediate spin parameters. \label{fig:lambda}

%As discussed in \citet{Hallenbeck2014}, 

\section{Discussion}

\label{sec:discussion}
%\citet{Hallenbeck2014} examined the \HI\ content of two other HIghMass galaxies: UGC 9037 and UGC 12506. In brief, UGC 9037 was observed to have \HI\ surface densities above the 10\solarmassespersquareparsec\ saturation threshold, a moderately unstable disk according to its Toomre $Q$, non-circular flows an, and a centrally peaked star formation rate. Its $\lambda=0.07$ is also slightly higher than average; we shall refer to this galaxy as the ``moderately high'' spin parameter galaxy. The authors of that work argue that the best explanation of its current state was that it was in transition, with gas flows caused by its weak bar and nearby companions. Left unresolved was whether this state was temporary, or quasi-static, with unobserved fresh gas from the cosmic web sustaining it. UGC 12506 had complementary low \HI\ surface densities, appeared to be stable across the disk, and has a $\lambda=0.15$. Optically, the galaxy is bluer and has a low surface brightness, properties which are believed to be correlated with high spin parameters. We shall thus refer to UGC 12506 as the ``high'' spin parameter galaxy.

%It is in the light of these previous results that we consider UGC 6168, 7899, and NGC 5230.

%\subsection{UGC 6168 and UGC 7899}

\noindent
UGC 9037 and 12506---the two HIghMass galaxies previously studied in detail by \citet{Hallenbeck2014}---were found to have very different properties. UGC 12506 has low surface densities of \HI\ (typically $1-5$\solarmassespersquareparsec\ at radii from $10-40$ kpc), and is an LSB. These properties can all be explained by its very high dark matter halo spin parameter ($\lambda=0.15$). UGC 9037, on the other hand, has an above-average but unexceptional spin parameter ($\lambda=0.07$). Its \HI\ has high ($>10 \solarmassespersquareparsec$) surface densities at $r<10$ kpc, and correspondingly the Toomre Q at most radii was moderately unstable, especially in comparison with the stable \HI\ disk of UGC 12506. UGC 9037 also has high-velocity gas inflows at all radii. UGC 9037's above average spin parameter may have suppressed star formation over much of cosmic history. However, the high surface densities of \HI\ and inflowing gas suggest a recent enhancement in star formation in comparison with its time-averaged rate. We thus see two very different states for these two galaxies: UGC 12506 remains in a low surface density, suppressed star formation state, while UGC 9037 is beginning a phase of enhanced star formation---possibly triggered by recently acquired gas.

%These findings are in agreement with \citet{Huang2014}, who came to this conclusion based on studying the \Halpha\ emission of the HIghMass galaxies. We thus call UGC 9037 a galaxy in transition.

Overall, the three galaxies discussed in this work (UGC 6168, UGC 7899, and NGC 5230) appear more like UGC 9037 than UGC 12506. First, none of the three are LSB galaxies. They all have typical star formation rates for their \Htwo\ masses, and have short star formation efficiency scale lengths---that is, their star formation quickly becomes extremely inefficient where the interstellar medium is \HI-dominated in comparison with the THINGS sample. These properties strongly suggest that any possible star formation bottleneck is in the \HI\ to \Htwo\ conversion, and not in the conversion of \Htwo\ to \HI. In addition, for the two galaxies which a spin parameter can be measured (UGC 6168 and UGC 7899), $\lambda$ is found to be above average, but not exceptionally so, and gas surface densities are found to reach typical values ($\sim10 \solarmassespersquareparsec$) over a range of radii. This is in contrast with the theoretical prediction that higher spin parameters are theoretically associated with lower gas and star formation surface densities (e.g. \citealt{Boissier2000}). We thus claim that the galaxies in this work are transitioning from a long history of suppressed star formation to a more active phase. These findings are in agreement with \citet{Huang2014}, who came to the same conclusion based on studying the \Halpha\ emission of the HIghMass galaxies.
%Overall, we thus claim that the the galaxies discussed in this work are galaxies in transition from a previous, like UGC 9037---consistent with possessing moderately high spin parameters, and being in transition.

UGC 6168, like UGC 9037 is observed to have a moderately unstable gas disk across a wide range of radii. It is also possible that UGC 6168 has inflowing gas, but the observed noncircular flows in the galaxy are of only marginal ($1-2\sigma$) significance. The strongest indication of noncircular flows is the misalignment between the average position angle of the \HI\ and CO gas phases. It does not have the high \HI\ surface densities observed in UGC 9037, instead saturating at the typical $10 \solarmassespersquareparsec$. Finally, we have calculated its spin parameter to be $\lambda=0.09$, an above average, but not extremely large value.

UGC 7899 shares many properties with UGC 6168 and UGC 9037: it is not an LSB, it has a moderately unstable disk, and above-average $\lambda$. Like the previously studied UGC 9037, the \HI\ in the center of the galaxy is not depleted, but saturates at $10 \solarmassespersquareparsec$; we also see high surface densities of \Htwo. Unlike UGC 6168 and UGC 7899, there is no evidence of noncircular motion in the gas disk of the galaxy. It is unique among the HIghMass galaxies so far presented in that its disk shows some warping, which could be indicative of recent cold accretion from the intergalactic medium. However, warps in \HI\ disks at large radii have long been observed to be common, even in relatively isolated galaxies (\citealt{Sancisi1976}; \citealt{Bosma1981}; \citealt{Briggs1990}; \citealt{vanderKruit2011}).

Because we are unable to make a clear case for the inclination---and thus the surface densities or dark matter profile---the case for NGC 5230 is more difficult. It is easiest to compare it with the other galaxies in this work: despite high \HI\ gas masses and GFs, all three show typical values of SFE(\HI) and SFE(gas) in comparison with other \HI-selected galaxies and the optically-selected GASS sample, respectively. All three show a shorter SFE length scale than for the local spirals of THINGS. In addition, all three show evidence for a much smaller $R_\text{HI}$ than is expected for their \HI\ masses, regardless of what inclination is assumed for NGC 5230.

\begin{figure}
\begin{center}
 \epsscale{1.0}
	\plotone{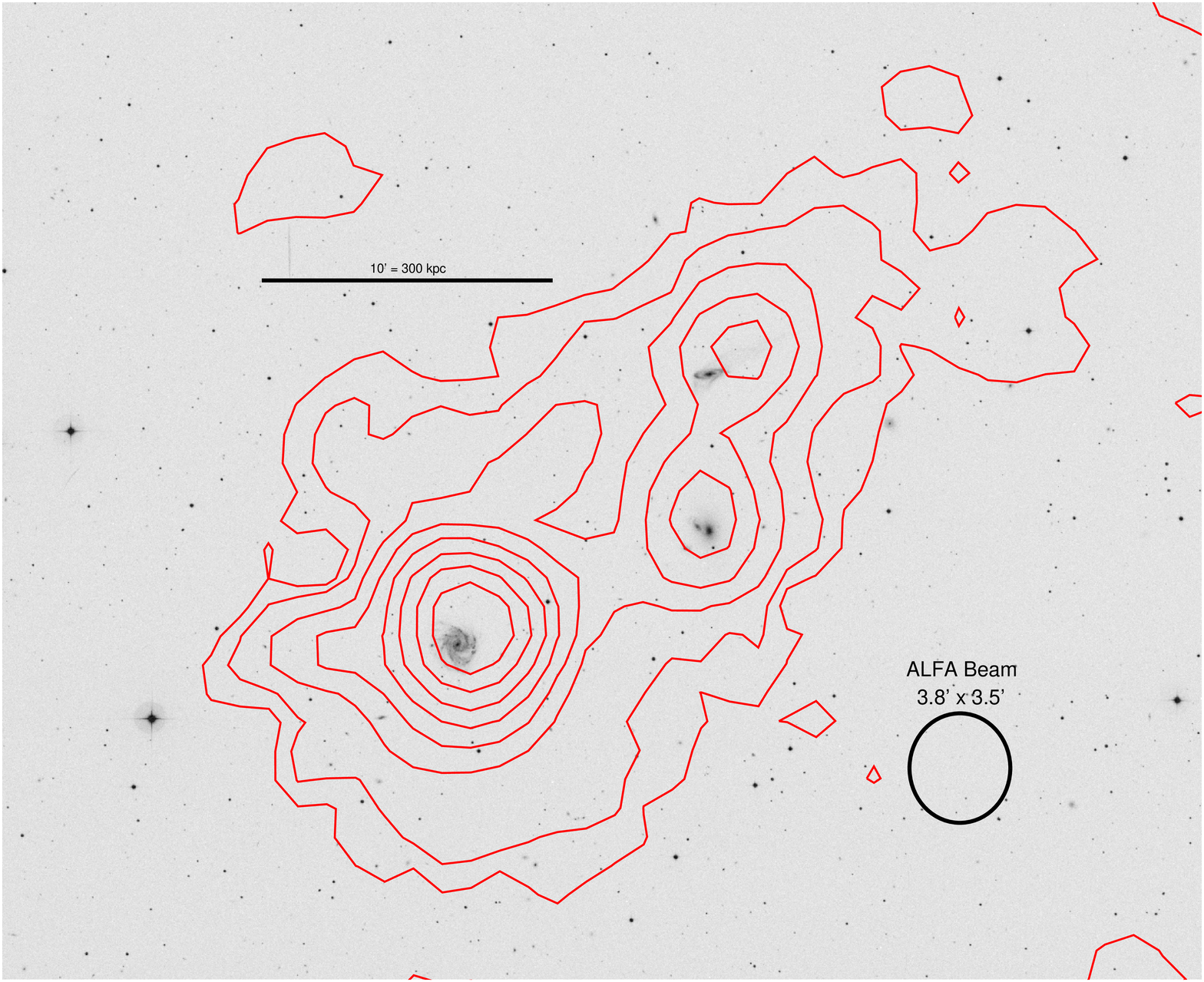}
\end{center}
\caption{Optical (DSS II/Blue) image of NGC 5230 (east) and its neighbors NGC 5222 (elliptical galaxy; west) and NGC 5221 (irregular spiral; northwest). Overlaid are ALFALFA contours, starting from $0.1$ Jy km s$^{-1}$ beam$^{-1}$ ($3\sigma$) and increasing by a factor of $\sqrt{2}$ at each additional contour. The peaks are roughly 10 and 3 Jy km s$^{-1}$ beam$^{-1}$ for NGC 5230 and each of its neighbors, respectively. Several tidal tails and bridges are clearly visible outside the optical galaxies, as well as a significant amount of low column density gas distributed throughout the group. The apparent tail from NGC 5230 pointing to the east is most likely an artifact from the ALFA beam and the ALFALFA grids, and not a real feature.\label{fig:group-gas}}
\end{figure}

NGC 5230's neighbors may be the most important clues to understanding the galaxy. There are two galaxies of similar size within 1 Mpc of NGC 5230: NGC 5222, at a projected distance of 300 kpc to the west, and NGC 5221, 400 kpc to the northwest. NGC 5222 is an elliptical galaxy hosting an AGN, and has an optically much smaller blue companion. More interesting is NGC 5221, an irregular spiral with a long tail pointing to the northwest. %Optically (Fig. \ref{fig:images-UGC8573}), NGC 5230 does not appear to have recently interacted with well-developed spiral arms. The only irregularity is in the galaxy's southern spiral arm, which is both less tightly wound and traceable out to greater radii than the northern arms. The southern arm also has several clear star formation knots near its western tip, which the other arms lack.
An optical image of NGC 5230 and its neighbors can be found in Figure \ref{fig:group-gas}, with contours from ALFALFA overlaid. These three galaxies are embedded in a common \HI\ envelope with a significant amount of gas: both NGC 5222's blue companion and NGC 5221 are detected in ALFALFA with $\log M_\text{HI}=9.96$ and $10.02$, respectively. A number of tidal tails and otherwise extragalactic gas can be observed in the ALFALFA data cubes. A bridge or between NGC 5230 and NGC 5221 is visible, and tails between NGC 5222 and its neighbors are possible, but unresolved due to the large ($\sim120$ kpc at 88 Mpc) ALFALFA beam. It is thus a strong possibility that NGC 5230's current state is due to its clear interaction with its neighbors. Its gas is likely compressed due to tidal torques, yet its star formation efficiency can remain low because much of the gas nominally associated with it is extended and at low column density.%In short, it is clear that these galaxies have interacted significantly in the recent past.

\section{Conclusions}
\label{sec:conclusions}
\noindent
The HIghMass sample is a selection of 34 galaxies from the ALFALFA 40\% data release which all have high \HI\ masses and are all gas rich for their stellar masses. We have presented resolved \HI\ and \Htwo\ studies of three HIghMass galaxies, UGC 6168, UGC 7899, and NGC 5230. Along with UGC 9037 and UGC 12506, this brings the total number of resolved gas studies of the HIghMass galaxies to five. None of the galaxies in this work appear to host extremely high dark matter spin parameters like UGC 12506. Instead, most of the galaxies so far appear more like the previously-studied UGC 9037: galaxies in transition from a gas-rich but inactive phase to a phase of active star formation. %Additionally, of the five galaxies observed, only one (UGC 7899) has a significant warp in the \HI\ disk, as would be expected if a significant mass of \HI\ had been recently accreted.
The galaxies in this work display the following properties:
\begin{itemize}
\item \textbf{High HI masses and high gas fractions} compared with an optically selected sample, like all HIghMass galaxies.
%\item \textbf{Color gradient reversal}, with the inner and outer radii of the galaxies relatively red in the optical compared with the relatively blue intermediate radii, as discussed in \citet{Huang2014}.
\item \textbf{More concentrated HI disks} are observed than are expected for their \HI\ masses. This was not observed in either of the HIghMass galaxies previously studied.
%\item \textbf{Significant non-circular velocities} over a $>10$ kpc range of radii; this property is shared with the moderate spin HIghMass galaxy. Because the \HI\ disks are concentrated, we infer that these non-circular velocities are likely inflows of gas, rather than outflows.
\item \textbf{Moderately unstable disks,} with values of both Toomre $Q$ and the \QN\ of \citet{RomeoFalstad2013} $\lesssim2$ over a wide range of radii, are observed in UGC 6168 and UGC 7899.
\item \textbf{Typical \Htwo\ star formation efficiencies} compared with two optically selected samples: the local spiral galaxies of THINGS and COLD-GASS galaxies of similar stellar mass. This comparison holds for both global (both samples) or resolved SFEs (THINGS).
\item \textbf{Mixed HI star formation efficiencies}. Globally, their SFE(\HI) is typical for an \HI-selected sample, but in the \HI-dominated region of each galaxy, SFE as a function of radius declines with a shorter scale length ($0.19R_{25}$) than the spirals of the THINGS sample ($0.25R_{25}$).
\item \textbf{Above average spin parameters} compared with an optically selected sample are observed for UGC 6168 and UGC 7899 ($\lambda = 0.09$ and $0.08$); we cannot directly measure $\lambda$ for NGC 5230.
\end{itemize}
Individually, there are a few unique features in each galaxy, which hint at their past, and why their \HI\ content is so large for their stellar masses: 
\begin{itemize}
\item For \textbf{UGC 6168}, the average position angles in the \HI\ and \Htwo\ phases do not match, which indicates non-circular motions in its gas disk.
\item \textbf{UGC 7899} is the only HIghMass galaxy thus far for which a warp has been observed in the outer \HI\ disk, which may be indicative of accretion of cold gas from the intergalactic medium.
\item \textbf{NGC 5230} has two lower-mass neighbors, all sharing a common \HI\ envelope. Its high \HI\ mass but typical star formation rate may be partially explained by gas taken from its neighbors which has not yet settled into the galactic disk.
\end{itemize}

%We further split the galaxies explored into two subclasses. First, UGC 6168 and UGC 7899 have predicted $\lambda=0.08-0.09$, which make both overall very similar to UGC 9037. However, neither has any nearby companions and only UGC 6168 has a hint of a bar. This suggests that secular processes may be much more important than the group environment for UGC 9037. Finally, NGC 5230 appears to have recently interacted with its neighbors NGC 5222 and NGC 5221. Tidal torques removing angular momentum 

%Most interestingly, it is possible for many HIghMass galaxies, the low local SFE(\HI) may be due to high inflow rates. If we instead consider the condition of the gas at the onset of recent star formation (100 My ago), then SFE(\HI) is normal as out to the largest radii that star formation has been traced. Future resolved studies will determine whether this explanation is consistent for more of the HIghMass galaxies.

\section*{Acknowledgements}
\noindent
This work has been supported by NSF-AST-0606007 and AST-1107390, NASA/JPL Spitzer RSA/73350, grants from the Brinson Foundation, and a Student Observing Support award from NRAO.\\
\\
This work is based in part on observations made with the Combined Array for Research in Millimeter-wavelength Astronomy (CARMA). Support for CARMA construction was derived from the Gordon and Betty Moore Foundation, the Kenneth T. and Eileen L. Norris Foundation, the James S. McDonnell Foundation, the Associates of the California Institute of Technology, the University of Chicago, the states of California, Illinois, and Maryland, and the National Science Foundation. CARMA development and operations was supported by the National Science Foundation under a cooperative agreement, and by the CARMA partner universities.\\
\\
This work is based in part on observations made with the Karl G. Jansky Very Large Array, a facility of the National Radio Astronomy Observatory (NRAO). The NRAO is a facility of the National Science Foundation operated under cooperative agreement by Associated Universities, Inc.
\\
This work is based in part on observations made with the Arecibo Observatory. The Arecibo Observatory is operated by SRI International under a cooperative agreement with the National Science Foundation (AST-1100968), and in alliance with Ana G. M\'{e}ndez-Universidad Metropolitana, and the Universities Space Research Association.
\\
This work is based in part on observations made with the Spitzer Space Telescope, which is operated by the Jet Propulsion Laboratory, California Institute of Technology under a contract with NASA.\\
\\
This work has made use of THINGS, `The \HI\ Nearby Galaxy Survey' \citealt{THINGS}.

\end{document}